\documentclass[aps,pra,twocolumn,groupedaddress,floatfix, nofootinbib]{revtex4-2}
\usepackage[T1]{fontenc}
\usepackage{graphicx}
\usepackage{amsmath}
\usepackage{bm}
\usepackage{braket}
\usepackage{eqnarray}
\usepackage{dsfont}
\usepackage{multirow}

\newcommand\sqnorm{\frac{1}{\sqrt{2}}}
\newcommand\refer[1]{(\ref{#1})}

\begin{document}

\title{Quantum repeaters based on stationary Gottesman-Kitaev-Preskill qubits}
\author{Stefan H{\"a}ussler and Peter van Loock}
\affiliation{Institute of Physics, Johannes Gutenberg-Universit{\"a}t Mainz, Staudingerweg 7, 55128 Mainz, Germany}
\date{\today}

\begin{abstract}
Quantum repeaters that incorporate quantum error correction codes have been shown to be a promising alternative compared with the original quantum repeaters that rely upon probabilistic quantum error detection depending on classical communication over remote repeater stations. A particularly efficient way of encoding qubits into an error correction code is through bosonic codes where even a single oscillator mode serves as a sufficiently large, physical system. Here we consider the bosonic Gottesman-Kitaev-Preskill (GKP) code as a natural choice for a loss-correction-based quantum repeater. However, unlike existing treatments, we focus on the excitation loss that occurs in the local, stationary memory qubits as represented by, for instance, collective atomic spin modes. We analyze and assess the performance of such a GKP-based quantum repeater where, apart from the initial state generations and distributions, all operations can be performed via deterministic linear mode transformations, as opposed to other existing memory-based quantum repeater schemes.                   
\end{abstract}

\maketitle

\section{Introduction}
Entangled quantum states distributed over long distances constitute a crucial requirement for many possible quantum technologies, such as quantum key distribution (QKD) or distributed quantum computing \cite{distributed, eco}. However, the most obvious way of obtaining such states, namely to entangle two qubits locally and subsequently transmit at least one of them to a remote receiver, is far from optimal due to transmission losses. Typically, these losses scale exponentially with distance $L$ and create an upper bound for the achievable transmission rates \cite{Takeoka2014}. In the most relevant case of photonic qubits in a fiber optic cable, this bound, referred to as the PLOB bound \cite{plob}, is mathematically expressed as $-\log_2(1 - e^{-L/L_\text{att}})\approx 1.44\, e^{-L/L_\text{att}} $, where $L_\text{att} = 22$km denotes the characteristic attenuation length of the fiber and $L$ must not be too small for the approximation to hold. 

To achieve rates above the PLOB bound, intermediate stations placed along the communication channel and acting as quantum repeaters, or equivalently turning the whole system into what is called a quantum repeater, are necessary modifications. It has become useful to classify quantum repeaters into three generations by the methods of dealing with transmission loss and operational errors \cite{generations}, although the basic concept of partitioning the total distance into smaller segments of length $L_0$ is shared by all of them. 

In repeaters of the first and second generations, entangled states are distributed in each segment with a success probability scaling as $\exp(-L_0/L_\text{att})$, and successively transformed into remote entangled states via entanglement swapping. A necessary requirement for such repeaters is the availability of quantum memories, typically in the form of stationary matter qubits. Otherwise, without the possibility of storing successfully distributed pairs, a complete distribution event over the entire distance would depend on waiting until all segments happen to be successful simultaneously to be connected, which occurs, at best, with probability 
$[e^{-L_0/L_\text{att}}]^n = e^{-L/L_\text{att}}$ for $n=L/L_0$ segments, thus reproducing the point-to-point performance. 

The essential difference between the first and second generations consists in how imperfections and errors of the quantum memory qubits are addressed: for this purpose, the first generation uses entanglement distillation to detect and partially eliminate such errors, including (two-way) classical communication performed even over the total distance $L$, whereas the second generation employs quantum error correction (QEC) to not only detect but also correct errors that occur on the stationary qubits. This allows the second generation to limit any (two-way) classical communication to distances of the order of $L_0$ (provided the entanglement swapping can be done deterministically), which means that for $L_0\sim 100{\rm m}$ the repeater clock rates, as determined by the classical signalling, approach values near the local, inverse light-matter interaction times (typically, $\sim$MHz). The errors that occur at the many faulty stations are then dealt with by the QEC code for the stationary qubits. 

Finally, the third generation relies upon QEC for both transmission and operational errors, and so dispenses with memories and extra waiting times for classical signals altogether. In principle, the third generation can achieve the highest rates (per second), however, it also imposes the highest demands on optical channel loss correction and gate fidelities, and is thus challenging to implement in practice. More specifically, it can be operated, in principle, at the same clock rate as a point-to-point link (typically, $\sim$GHz), but it requires a higher frequency of repeater stations (typically, one station at every few km) and, in particular, does not permit channel loss as high as 50\% in a single segment. As a consequence, it appears that second-generation quantum repeaters are a good compromise between reasonably practical clock rates and experimental feasibility.    

In this work, we focus on the second repeater generation, where the QEC is achieved with the help of the bosonic, Gottesman-Kitaev-Preskill (GKP) code \cite{gkp}.  
As the quantum memories that can store GKP states, we envision atomic spin ensembles, i.e., a collection of $N$ two-level systems as described, e.g., in Refs.~\cite{hammerer} and \cite{lightgkp}. The possibility of defining GKP-like states in such an ensemble even at moderately small $N$ has been discussed in detail in Ref.~\cite{ensgkp}; however, here we assume sufficiently large ensembles in order to be able to describe the atomic GKP states as they are known from continuous-variable (CV) applications. 

More precisely, the ensemble can be represented as a collectively acting spin wave whose ``magnonic'' excitations behave as bosons. Memory loss in this case corresponds to excitation loss, where the bosonic system is subject to a collective bosonic loss channel locally acting on the spin memories. Thus, the system can be, in principle, protected by a bosonic QEC code, for which the GKP code is a prominent choice. In particular, the GKP code allows for a deterministic implementation of all logical Clifford gates on the qubits via physical Gaussian operations. Clifford gates are all that is needed in a quantum repeater based on QEC codes. Thus, the error correction and especially also the entanglement swapping can be performed deterministically with stationary GKP qubits - in principle, even with only linear (Gaussian) bosonic mode transformations, provided the initial states can be prepared in the first place \cite{konno24, Fluhmann2019-jv, Campagne-Ibarcq2020}.        
Note that other proposals for memoryless, in principle all-optical, third-generation quantum repeaters based on optical QKP qubits already exist \cite{gkp3rdgen1, alloptical, gkp3rdgen2, frank_quditgkprepeater}.

The paper is structured as follows.
In Sections~\ref{subsec::ESdis} to \ref{subsec::ESqubit}, we shall discuss the building blocks of our quantum repeater protocol, i.e., the initial entanglement distribution, the GKP error correction, and the necessary amplification (or alternative) techniques to transform loss errors into Gaussian shifts, as well as the deterministic, Gaussian-operation-based entanglement swapping for GKP-encoded qubits. In Sections~\ref{subsec::ensmemory} and \ref{subsec::ensoperations}, we briefly review the formalism of atomic spin ensembles and the Holstein-Primakoff approximation, which allows to treat a subset of the ensemble state space as an approximate bosonic phase space, as well as the possibilities of implementing operations like amplification and entanglement swapping on modes stored in an atomic ensemble memory. In Section~\ref{sec::rateanalysis}, we will then quantify and assess the performance of our quantum repeater based on the secret key rate in quantum key distribution (QKD) as a figure of merit.

\section{Quantum repeaters based on stationary atomic GKP qubits}
In this section, the functioning principle of a ``second-generation'' quantum repeater \cite{generations} shall be discussed in more detail, with a particular focus on the implementation of the individual steps for GKP-encoded qubits. 
We will first discuss the entanglement distribution, then the concept of GKP encoding and quantum error correction adapted to our scheme based on atomic ensembles, and finally, how to realize entanglement swapping and error syndrome detection at the same time with the atomic ensemble memories.  
\begin{figure}
\includegraphics[width=0.45\textwidth]{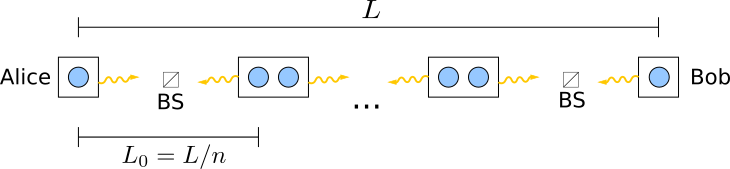}
\caption{Schematic illustration of a memory-based quantum repeater: the total distance $L$ is partitioned into $n$ segments of length $L_0$ by inserting intermediate stations. Every station contains two quantum memories to enable entanglement swapping between neighboring segments. In a ``second-generation'' quantum repeater, these stationary qubits are logical qubits protected against memory (storage and gate) errors by a suitable quantum error correction code. The encoded entanglement distribution is achieved by locally entangling photonic qubits with stationary GKP qubits, storing the latter, and sending, combining at a beam splitter (BS), and measuring the former in the middle of each segment.}
\label{fig::Intro:setup}
\end{figure}

\subsection{Entanglement distribution}
\label{subsec::ESdis}
The first step in a repeater protocol, and requirement for all following steps like entanglement swapping, is the distribution of entangled states across the length of a single segment (see Fig.~1). This work focuses on repeaters with GKP-encoded qubits, which means that an entangled Bell state of the form $\sqnorm(\ket{\overline{0}}\ket{\overline{1}} \pm \ket{\overline{1}}\ket{\overline{0}})$ with the logical GKP states $\ket{\overline{0}}$ and  $\ket{\overline{1}}$ should be created in the collective memory modes connected by the segment. The concept of GKP qubits will be introduced in the subsequent section. 

There are, in principle, at least two options to achieve this. On the one hand, the Bell-state GKP qubit pair could be created locally, for instance, in the middle of the segment with a suitable optical source and state generator, and subsequently transmitted to the memory stations where it is written into the quantum memories. On the other hand, non-entangled, individual GKP qubits can also be created directly in the memories and later be entangled using other means. Generally, an approach where the photonic qubits sent through the fiber channel are simple single-photon-based qubits rather than (brighter, multi-photon) GKP qubits is more consistent with the notion of a second-generation quantum repeater, where QEC is only used to mitigate memory loss (or generally, local errors) but not loss arising in transmission. 
In the following, we will illustrate the simplest scenario where single-photon-based qubits are combined in the middle of each repeater segment for GKP Bell pair distribution.   

Let us assume that it is possible to locally create entangled states of the form 
\begin{equation}
\sqnorm(\ket{\overline{0}}\ket{H} + \ket{\overline{1}}\ket{V})
\end{equation}
between a stationary GKP qubit and a photonic dual-rail qubit, where $\ket{H}$ and $\ket{V}$ denote the respective single-photon polarization states of the photonic mode. A possible way of achieving this, in principle, is to employ a CV cubic gate of the form $e^{i \sqrt{\pi} \hat p \otimes \hat n }$ that acts upon a bosonic spin mode initialized in a GKP state $\ket{\overline{0}}$ together with one of the two polarization modes of the optical dual-rail qubit, say the vertically polarized mode, while the photonic qubit starts in a superposition state $\sqnorm(\ket{H} + \ket{V})\equiv \sqnorm(\ket{10} + \ket{01})$. Since the gate results in a photon-number-dependent position shift in the phase space of the spin mode by $\sqrt{\pi}$, one obtains $e^{i \sqrt{\pi} \hat p_1 \otimes \hat n_3 } (\ket{\overline{0}} \otimes \sqnorm(\ket{10} + \ket{01}) = \sqnorm(\ket{\overline{0}}\ket{H} + \ket{\overline{1}}\ket{V})$, where the operator indices denote the modes -- the spin and the two polarization modes, respectively. Here, the GKP state $\ket{\overline{1}}$ is position-shifted by $\sqrt{\pi}$ compared to $\ket{\overline{0}}$, as discussed in more detail in the subsequent section. While an obvious experimental complication of this approach is that it is difficult to implement such a non-Gaussian interaction gate upon a free-space atomic ensemble and a photonic qubit, possible solutions may be to employ interacting atoms and also to place them in cavities \cite{remperydbergexp}  
or to let the free-space ensembles interact with the light fields for sufficiently long \cite{lightmed}.
Note that single-mode cubic phase gates \cite{gkp, akira_nl_squeeze}
and passive linear mode transformations are sufficient to
create GKP states and to realize the above controlled two-mode gates \cite{niklas_cpg}. 

After the initial preparation of the light-atom states $\sqnorm(\ket{\overline{0}}\ket{H} + \ket{\overline{1}}\ket{V})$ in all repeater stations, the photons are sent through an optical fiber channel towards the middle of each segment, while the GKP qubits remain in the memories at the stations. 
The photons arriving from both directions at the midpoint between the stations then pass through a beam splitter (Fig.~1) that transforms the creation mode operators as
\begin{equation}
\begin{pmatrix}a_i^{\dagger\prime}\\ b_i^{\dagger\prime}\end{pmatrix} = 
\sqnorm\begin{pmatrix}1 & 1\\-1& 1\end{pmatrix}\begin{pmatrix}a_i^\dagger \\ b_i^\dagger\end{pmatrix}.
\end{equation}
Here, $a$ and $b$ are the spatial propagation directions and input mode operators, and $a^\prime$ and $b^\prime$ the operators of the beam splitter's output modes, while $i \in \{H, V\}$ denotes the polarization state (mode) for each spatial mode. After the beam splitter, the number of photons of a certain polarization is counted in each output mode using polarizing beam splitters followed by photon detectors. 

In principle, the four photon detectors can generate ten different count results for two incident photons, namely 2000, 0200, 0020, 0002, 1100, 1010, 1001, 0110, 0101 and 0011, where numbers represent photons counted in the modes $a_H^{\dagger\prime}$, $a_V^{\dagger\prime}$, $b_H^{\dagger\prime}$ and $b_V^{\dagger\prime}$ in this order. However, the two results 1010 and 0101 are not observed, because, as the setup does not change polarizations, measuring two photons of equal polarization implies incident photons of equal polarization, and it is well known that in this case both photons are always observed at the same beam splitter output port \cite{hongoumandel}. 
The remaining results all occur with probability $\frac18$, but a Bell state in the GKP modes is not obtained when a ``bunched'' photon pattern is measured and it only emerges from observing two photons at different detectors, in which case the ensemble state is given by
\begin{equation}
\sqnorm  (\ket{\overline{1}\overline{0}} + \ket{\overline{0}\overline{1}})\,,
\end{equation}
for the measurement results 1100 and 0011, and by
\begin{equation}
\sqnorm  (\ket{\overline{1}\overline{0}} - \ket{\overline{0}\overline{1}})\,,
\end{equation}
for measurement results 1001 and 0110.

The entanglement distribution is therefore not deterministic, but probabilistic with a certain success probability, which is bounded from above by $\frac12$ for the procedure presented here. This value, along with other factors that can influence the success of the distribution, e.g. arising from the creation of the required states as well as light fiber coupling and frequency conversion efficiencies, is absorbed into a parameter $p_\text{link}$. 

The most important influence on the distribution success probability in a long-distance communication scenario, that is accounted for separately, is the length of each segment, since it is increasingly likely that photons do not reach the beam splitter due to a fiber loss (attenuation) probability growing exponentially with distance.
Overall, the entanglement distribution in one segment is successful with probability 
\begin{equation}
p = p_\text{link}\exp\left(-\frac{L_0}{L_\text{att}}\right),
\end{equation} 
and the statistical number of attempts until success in a single segment is distributed geometrically with an elementary probability $p$. 

\subsection{GKP quantum error correction}
\label{subsec::AS}
In our repeater scheme, we assume that entanglement swapping takes place as soon as possible.
Nevertheless, since entanglement distribution in each segment is a stochastic process, it can happen that the modes of a successfully distributed Bell pair need to wait in memory for a neighboring segment to also achieve successful distribution, being subject to memory loss in the meantime. The assumption of memory loss as the dominating memory imperfection is specific to our scheme based on stationary collective bosonic spin modes, in contrast to schemes that rely upon single-spin memory qubits where memory dephasing is the most common storage error model \cite{rateanalysis}.  
The memory loss in our scheme is modeled as a channel of the form
\begin{equation}
\label{eq::AS:losschannel}
\mathcal{L}(\eta)[\rho] = \text{Tr}_E[B(\eta)(\rho \otimes \rho_E)B^\dagger(\eta)],
\end{equation}
with the beam-splitter unitary $B(\eta)$ and the environment in the vacuum state $\rho_E = \ket{0}\bra{0}$. As shown, e.g. in Ref.~\cite{channels}, 
this loss channel can be converted into a Gaussian shift channel,
\begin{equation}
\mathcal{E}(\sigma^2)[\rho] = \frac{1}{\pi\sigma^2}\int d^2\beta\, D(\beta)\rho D^\dagger(\beta) \exp\left[-\frac{|\beta|^2}{\sigma^2}\right]\,,
\end{equation}
with variance $\sigma^2$ by applying an amplification channel defined analogously to Eq.~(\ref{eq::AS:losschannel}) but with a two-mode squeezing unitary replacing the beam splitter, and subsequently be corrected using GKP error correction as described later. Here, $D(\beta)$ is the quantum optical displacement operator.

The loss channel's amplitude depends on the waiting time and can be modeled as an exponential decay $\eta(t) = \exp(-\alpha^\prime t)$ with decay constant $\alpha^\prime$. For the analysis in later sections of this work it will be more useful not to consider time a continuous variable, but to measure it in discrete steps of duration $\tau = \frac{L_0}{c}$, where $c$ is the speed of light in the optical fiber, which is reduced by a factor of $\frac23$ compared to the vacuum speed of light. 

If we denote the number of time steps by $M$, the loss amplitude reads 
\begin{equation}
\eta(M) = \exp(-\alpha^\prime M\tau) = \exp(-\alpha M),
\end{equation}
with the new dimensionless decay constant $\alpha$ related to the memory coherence time $t_\text{coh}$ via $\alpha = \frac{\tau}{t_\text{coh}}$ (this parameter $\alpha$ plays the role of an inverse effective memory coherence time \cite{rateanalysis}, however, here with regards to memory loss rather than memory dephasing. For an overview of all the parameters used in the repeater analysis, see App.~\ref{listofsymbols}). 

It is a well-known fact that when converting losses to shifts, preamplification leads to a lower variance of the resulting Gaussian shift than postamplification, and therefore should be preferred. However, preamplification is not straightforward in the case of our repeater, as the waiting time $t_\text{wait}$ is a random variable and thus the amplitude of the loss channel is not known a priori. 
\begin{figure*}
\begin{minipage}{0.45\textwidth}
\includegraphics[width=\textwidth]{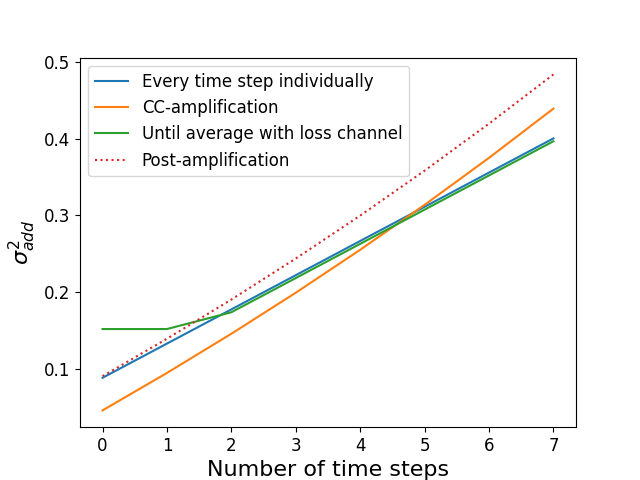}
\end{minipage}
\begin{minipage}{0.45\textwidth}
\includegraphics[width=\textwidth]{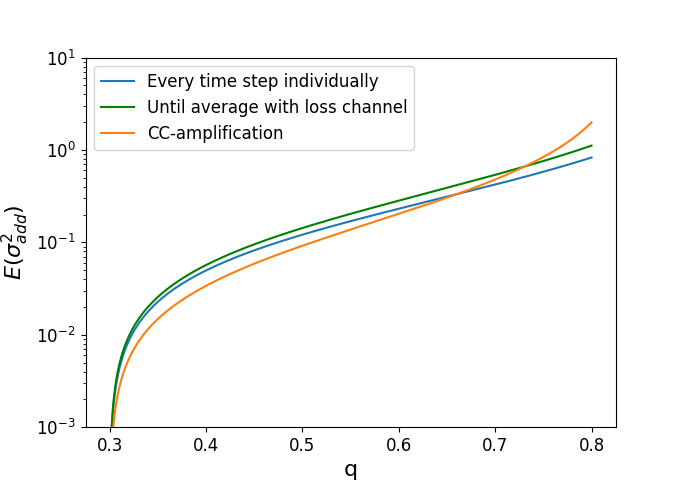}
\end{minipage}
\caption{Comparison of amplification strategies. Left: additional variance as a function of the number of time steps a spin mode has to wait for. Right: expectation value of the additional variance as a function of the distribution failure probability $q=1-p$. Both figures use $p_\text{link} = 0.7$, $t_\text{coh} = 1$ms.  }
\label{fig::AS:amplification}
\end{figure*}
A possible solution to this problem would be to preamplifiy for each time step individually, after that time step check whether a neighboring segment has succeeded and depending on the outcome either perform entanglement swapping or preamplify for the next time step. The variance incurred in this process in addition to the natural variance $\delta^2$ of the GKP states (as introduced below) is given by
\begin{equation}
\sigma^2_\text{add} = (t_\text{wait} + 2)(1 - e^{-\alpha}),
\end{equation}
as both modes are kept in memory for at least one time step, during which the dual-rail photon qubits propagate to the beam splitter and the information about the success of the Bell measurement is sent back to the stations, and the mode belonging to the segment finishing earlier further receives a variance of $1 - e^{-\alpha}$ from the preamplification in each time step spent waiting.  

Alternatively,  one could adjust the amplification of the first mode to the expectation value of $t_\text{wait}$ rounded to the closest integer, which we will abbreviate with $T$. This contributes $1 - e^{-(T+1)\alpha}$ to the variance. 
In case the real waiting time is greater than $T$, individual amplification for each step is used for the remaining time resulting in a linear increase similar to the first method, albeit starting at a lower variance.  

If the second segment finishes distribution faster than average, however, both modes still need to wait until time step $T$ in order to align the strength of loss and amplification before entanglement swapping can occur. This possibility of both modes waiting despite successful distribution constitutes the downside of this method. 

Overall, the additional variance acquired using this method amounts to
\begin{widetext}
\begin{equation}
\label{eq::AS:m2}
\sigma^2_\text{add} = \begin{cases}2 - e^{-(T+1)\alpha} - e^{-(T-t_\text{wait}+1)\alpha} & t_\text{wait} \leq T \\
1 - e^{-(T+1)\alpha} + (t_\text{wait}-T+1)(1 - e^{-\alpha}) & t_\text{wait} > T.
\end{cases}
\end{equation} 
\end{widetext}

The detrimental effect of both modes being ready for swapping but having to wait nonetheless can be somewhat mitigated by generating the necessary losses artificially in the form of a beam-splitter interaction with the environment instead of by waiting in memory. In the framework of atomic spin ensembles as quantum memories, such a loss channel can be applied by letting the ensemble interact with suitably polarized light modes. 

Thus, if the second segment finishes distribution before the average waiting time, a loss channel equivalent to the time difference to $T$ is applied to the ``older'' mode so as to compensate for the amplification having been too high, and creating a pure Gaussian shift. Equation (\ref{eq::AS:m2}) hence gets modified to
\begin{widetext}
\begin{equation}
\label{eq::AS:m2b}
\sigma^2_\text{add} = \begin{cases}2 - e^{-(T+1)\alpha} - e^{-\alpha} & t_\text{wait} \leq T \\
1 - e^{-(T+1)\alpha} + (t_\text{wait}-T+1)(1 - e^{-\alpha}) & t_\text{wait} > T,
\end{cases}
\end{equation}
\end{widetext}
from which it is apparent that variances could be reduced for the $ t_\text{wait} \leq T$ regime compared to the unmodified second method. 

Utilizing an artificial loss channel also enables the usage of a special amplification method as proposed in Ref.~\cite{alloptical}. A necessary condition for this method referred to as CC-amplification (where ``CC'' stands for ``classical computer'' as the amplification is done on the level of the classical postprocessing) is that the two modes on which the Bell measurement is performed later be subject to losses of equal strength. In our case, this means that the newly finished mode must undergo an artificial loss channel whose amplitude corresponds to the losses caused by waiting on the first mode. The additional variance is then given by
\begin{equation}
\sigma_\text{add}^2 = \frac{1-e^{-(t_\text{wait}+1)\alpha}}{e^{-(t_\text{wait}+1)\alpha}}.
\label{eq::AS:CC1}
\end{equation}

With the help of the probability distribution for $t_\text{wait}$ derived in App.~\ref{subsec::appGeo},  one can calculate the expected additional variance resulting from the different schemes. In the regime $\alpha T \ll 1$, valid for $q = 1 - p\lesssim 0.99$, one finds
\begin{equation}
\mathds{E}(\sigma^2_\text{add}) = (T + 2)\alpha
\end{equation}
for the first and
\begin{equation}
\label{eq::AS:av2b}
\mathds{E}(\sigma^2_\text{add}) = (T+2)\alpha + \frac{2\alpha q^{T+1}}{1-q^2}
\end{equation} 
for the modified second method. As the second summand in Eq.~(\ref{eq::AS:av2b}) is positive, the variance of the first method turns out less, rendering it the better method.

For the CC-amplification, the expectation value reads as (see App.~\ref{CCexpect}):
\begin{equation}
\label{eq::AS:CC}
\mathds{E}(\sigma_\text{add}^2) = \frac{p^2}{1-q^2}\left[\frac{1-e^{-\alpha}}{e^{-\alpha}} + \frac{2e^{2\alpha}q}{1-qe^{\alpha}} - \frac{2q}{1-q}\right].
\end{equation}\\

As apparent from Fig.~\ref{fig::AS:amplification}, CC-amplification performs best as long as the failure probability is not too high, i.e. the segment length $L_0$ is not too long. Numerically we find the thresholds listed in Table~\ref{tab::AS:l0s}. When performing the rate analysis, we assume CC-amplification whenever $L_0$ is shorter than the applicable threshold, and individual preamplification for each time step otherwise.  

\begin{table}
\begin{ruledtabular}
\begin{tabular}{cc|ccc}
 & & \multicolumn{3}{c}{$p_\text{link}$} \\
 & & 0.05 & 0.7 & 1\\
\hline
 \multirow[c]{3}{*}{$t_\text{coh}$}& 0.001s & 0.5km & 16km & 20km \\
 & 0.1s & 14km & 56km & 63km \\
 & 10s & 50km & 100km & 108km
 
\end{tabular}
\end{ruledtabular}
\caption{Thresholds for segment length $L_0$ such that CC-amplification results in a smaller variance than preamplification.}
\label{tab::AS:l0s}
\end{table}

Gaussian shifts can be detected using the GKP code \cite{gkp}, whose idealized basis states are defined as
\begin{subequations}
\begin{eqnarray}
\ket{\overline{0}} &= &\sum_{n \in \mathds{Z}}\ket{2n\sqrt{\pi}}_x\\
\ket{\overline{1}} &=  &\sum_{n \in \mathds{Z}}\ket{(2n + 1)\sqrt{\pi}}_x,
\end{eqnarray}
\end{subequations}
where kets with index $x$ denote $x$-quadrature eigenstates with $\hat{x}\ket{m}_x = m\ket{m}_x$.
\footnote{Incidentally, operators are usually written without the ``hat'' symbol in this paper, except when omitting it might lead to confusion.}
Ideal GKP states are thus infinite superpositions of quadrature eigenstates that can on their part be understood as displaced infinitely squeezed states. It is therefore obvious that  in practice, states of this form can never be realized exactly, but only be approximated by replacing delta-peaks in phase-space with narrow Gaussians and truncating the infinite sum on both sides. In theoretical descriptions, the truncation is often omitted and the infinite sum is instead superimposed by a second, wide Gaussian. 

Thus, realistic GKP states take the form
\begin{subequations}
\label{eq::realistic}
\begin{eqnarray}
\ket{\overline{0}}_\text{real} &\propto &\sum_{n\in\mathds{Z}}\exp\left[-\frac{(2n\sqrt{\pi})^2}{\Delta^2}\right]\\
&&\times\int dy\, \exp\left[-\frac{(2n\sqrt{\pi} - y)^2}{4\delta^2}\right]\ket{y}_x \nonumber\\
\ket{\overline{1}}_\text{real} &\propto &\sum_{n\in\mathds{Z}}\exp\left[-\frac{((2n + 1)\sqrt{\pi})^2}{\Delta^2}\right]\\
&&\times\int dy\, \exp\left[-\frac{((2n + 1)\sqrt{\pi} - y)^2}{4\delta^2}\right]\ket{y}_x \nonumber
\end{eqnarray} 
\end{subequations}
with the standard deviations $\Delta/\sqrt{2}$ and $\sqrt{2}\delta$ of the wide and narrow Gaussians respectively. The Wigner function of the realistic states is composed of two-dimensional Gaussians with variances $\sigma_x^2 = \delta^2$ and $\sigma_p^2 = \frac{1}{\Delta^2}$ arranged in a grid in phase space. In order to obtain symmetric variances in $x$ and $p$, it is common to set $\Delta = \frac{1}{\delta}$. \\

To simplify theoretical calculations, one often employs a Gaussian-noise approximation for the description of realistic GKP states, which means that the density operators of states as defined by Eq.~\refer{eq::realistic} are replaced with density operators of ideal states after the action of a Gaussian shift channel:
\begin{subequations}
\begin{eqnarray}
(\ket{\overline{0}}\bra{\overline{0}})_\text{real} & = &\mathcal{E}(\delta^2)\left[\ket{\overline{0}}\bra{\overline{0}}\right]\\
(\ket{\overline{1}}\bra{\overline{1}})_\text{real} & = &\mathcal{E}(\delta^2)\left[\ket{\overline{1}}\bra{\overline{1}}\right].
\end{eqnarray}
\end{subequations}
GKP states are well suited to detection and correction of random, normally distributed shifts in phase space, as long as the shifts' absolute value is not to large. One distinguishes two methods for finding the syndrome information. 

\begin{figure}
\includegraphics[width=0.45\textwidth]{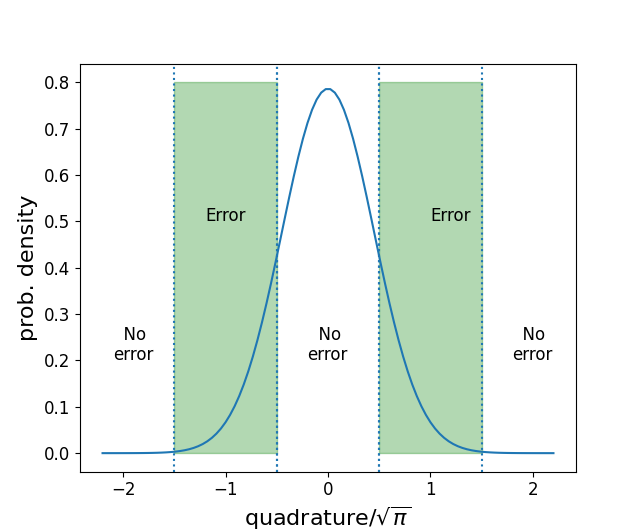}
\caption{Error correction of Gaussian shifts. The graph shows the distribution of the true shift, the dotted lines separate shaded regions where the syndrome interpretation leads to a logical error from those where no error occurs.}
\label{fig::GKP:fehler}
\end{figure}

The first method makes use of a SUM gate, in order to connect the data mode $A$ containing the GKP-encoded qubit with an ancillary mode $B$ initialized in the state $\ket{\overline{+}}$. After applying the SUM gate, the quadratures read
\begin{eqnarray}
x_A^\prime =& x_A, \hphantom{ + x_B} \qquad &p_A^\prime = p_A - p_B,\nonumber\\
x_B^\prime =& x_B + x_A, \qquad &p_B^\prime = p_B,
\end{eqnarray}
followed by a measurement of $x_B^\prime$, whose result we will denote $\overline{x_B^\prime}$. Since the ancillary mode can only contribute integer multiples of $\sqrt{\pi}$, any possible deviations must have been caused by errors in the data mode. Hence the shift of the data mode can be estimated by choosing that representative of the equivalence class $[\overline{x_B^\prime} \mod \sqrt{\pi}]$ that lies between $-\frac{\sqrt{\pi}}{2}$ and  $\frac{\sqrt{\pi}}{2}$. Sufficiently small shifts can thus be detected correctly. 

If, however, the absolute value of the true shift is larger than  $\frac{\sqrt{\pi}}{2}$, this method of interpretation will yield an incorrect value, resulting in the corrected state differing from the original state by a shift of $k\sqrt{\pi}$, $k\in \mathds{Z}$. For odd $k$ this results in a logical Pauli error, whereas for even $k$ theoretically no error occurs, as a shift by an even multiple of $\sqrt{\pi}$ maps the ideal code states onto themselves. This results in the striped pattern illustrated in Fig.~\ref{fig::GKP:fehler}. In the case of realistic states, the shifted and the original differ for even $k$ too, though, because the position of the maximum of the Gaussian envelope will have been shifted. 
If states are stored in a memory based on atomic ensembles this shifting can lead to the state's support leaving the region of phase-space where the Holstein-Primakoff approximation is valid. Errors of such magnitude are relatively unlikely, however. 

The second method of error correction using GKP states is based on quantum teleportation \cite{alloptical, knillstyle}. The data mode is mixed with one mode of a GKP-encoded Bell pair at a beam splitter and the output modes are subsequently measured using homodyne detection, which is equivalent to a measurement in the Bell basis, as explained below. The measurement result yields both the information necessary for teleportation, namely which operation to apply to the remaining mode in order to reproduce the original state, and the syndrome information for error correction, that is interpreted as in the first method.

\subsection{Entanglement swapping}
\label{subsec::ESqubit}
Entanglement Swapping \cite{teleport, swappingexp} aims at transforming two entangled qubit pairs with spatially separated components into one entangled pair consisting of the two remote qubits by performing a measurement in the Bell basis on the two central qubits, in order to increase the distance spanned by an entangled pair. The concrete physical implementation of the Bell measurement is dependent on the encoding of the qubits. 

The maximally entangled Bell states constituting the Bell basis of the two-qubit Hilbert space are defined as $\ket{\Phi^\pm} = \sqnorm(\ket{00} \pm \ket{11})$ and $\ket{\Psi^\pm} = \sqnorm(\ket{01} \pm \ket{10})$.
In analogy to Ref.~\cite{rateanalysis} we assume the initial state to consist of two $\ket{\Psi^+}$ states in the qubits $A$ and $C$, and $B$ and $D$ respectively, i.e. $\ket{\Psi^+}_{AC}\ket{\Psi^+}_{DB} $.
After the measurement, qubits $A$ and $B$ are in the same state that was measured on $C$ and $D$, where all measurement outcomes are equally likely. To recover the state $\ket{\Psi^+}$, the measurement result is transmitted classically to one of the qubits $A$ or $B$, where a Pauli operator is applied if necessary. 

In a quantum repeater where many swapping processes should occur successively or even simultaneously, it is not necessary to wait until the result arrives at the remote qubits of the newly formed segment via classical communication, which would require long transmission times particularly for later swappings when the distance is multiple times longer than the length of an original segment $L_0$. Instead the information is sent towards Alice or Bob, and the Bell state created by the measurement is immediately reused for further swappings without modification. Given that the time and relative place in the repeater chain for each swapping are transmitted along with the measurement result, the information about which Bell state is eventually created between Alice and Bob can be reconstructed at the end. 

In the case of GKP-encoded qubits the Bell measurement can be realized with the aid of a 50:50 beam splitter followed by homodyne detection. The beam splitter transforms the quadratures as
\begin{eqnarray}
x_A^\prime &= \sqnorm(x_A + x_B), \qquad &p_A^\prime = \sqnorm( p_A + p_B),\nonumber\\
x_B^\prime &= \sqnorm(x_B - x_A), \qquad &p_B^\prime = \sqnorm(p_B - p_A),
\end{eqnarray}
such that homodyning the output modes effectively measures the sum of the $x$-quadratures and the difference of the $p$-quadratures. 

In a $\Phi$ state, meaning $\ket{\Phi^+}$ or $\ket{\Phi^-}$, both modes are in the same basis state $\ket{\overline{0}}$ or $\ket{\overline{1}}$, although it is not determined in which one. Since $\ket{\overline{0}}$ can only contribute even and $\ket{\overline{1}}$ only odd multiples of $\sqrt{\pi}$ to the $x$-quadrature, the measurement outcome for the sum of $x$-quadratures divided by $\sqrt{\pi}$ must be either the sum of two even numbers or the sum of two odd numbers, yielding an even number overall. 
For $\Psi$ states the opposite holds: the two modes are in different basis states, so that one always measures the sum of an even and an odd number, yielding an odd number. Thus the measurement result $\overline{x_A^\prime}$ allows to distinguish between $\Phi$ and $\Psi$ states.  

Using the second measurement result $\overline{p_B^\prime}$ it is possible to distinguish $+$ and $-$ states. To see this, it is helpful to write the GKP states in momentum representation as
\begin{subequations}
\begin{eqnarray}
\ket{\overline{0}} &= &\sum_{n \in \mathds{Z}}\ket{n\sqrt{\pi}}_p\\
\ket{\overline{1}} &=  &\sum_{n \in \mathds{Z}}(-1)^n\ket{n\sqrt{\pi}}_p,
\end{eqnarray}
\end{subequations}
where we used the relation $\ket{x}_x = \frac{1}{\sqrt{2\pi}}\int dp\, e^{-ipx}\ket{p}_p$. The Bell states now take the form
\begin{subequations}
\begin{eqnarray}
\ket{\Phi^\pm} &\propto &\sum_{nm}\left(\ket{n\sqrt{\pi}}_p\ket{m\sqrt{\pi}}_p \right.\nonumber\\
&&\left.\pm (-1)^{m+n}\ket{n\sqrt{\pi}}_p\ket{m\sqrt{\pi}}_p\right)  \\
\ket{\Psi^\pm} &\propto &\sum_{nm}\left((-1)^m\ket{n\sqrt{\pi}}_p\ket{m\sqrt{\pi}}_p \right.\nonumber\\
&&\left.\pm (-1)^n\ket{n\sqrt{\pi}}_p\ket{m\sqrt{\pi}}_p\right).
\end{eqnarray}
\end{subequations}
It is easy to see that $\ket{\Phi^+}$ contains only terms with even $m+n$, whereas $\ket{\Phi^-}$ contains only terms with odd $m+n$. However, the difference of two integers has the same parity property as their sum, which means that for $\ket{\Phi^+}$ an even and for $\ket{\Phi^-}$ an odd difference of $p$-quadratures will be measured. 
For $\Psi$ states one finds that $\ket{\Psi^+}$ only contains terms where $m$ and $n$ have the same parity, while in all remaining terms of $\ket{\Psi^-}$ they have opposite parity. Since the difference of two integers of equal parity is even and that of two integers of opposite parity is odd, an even measurement result must correspond to a $+$ state and an odd one to a $-$ state. 

 Instead of a beam splitter a SUM gate can also be used, since it, too, replaces $p_A$ with the difference of momentum quadratures and $x_B$ with the sum of position quadratures, and the other quadratures are not measured. 
All in all the measurement yields exactly the information necessary to distinguish the four Bell states and therefore constitutes a Bell measurement.

\subsection{GKP code for atomic ensemble memories}
\label{subsec::ensmemory}
As the ability to store quantum states in memory is essential for a second generation repeater, we will now briefly discuss one possible realization of a quantum memory, in the form of an atomic spin ensemble. A more detailed discussion with a particular focus on the interaction of ensembles with light and the possibilities of applying operations like amplification and entanglement swapping to the stored states can be found in App.~\ref{ensemblephysics}.

For our quantum memory we are interested in the completely symmetric subspace of an ensemble of $N$ atoms modeled as two-level systems with the orthogonal states $\ket{1}$ and $\ket{2}$. This subspace is spanned by states of the form
\begin{equation}
\ket{k} = \frac{(b^\dagger)^{N-k}}{\sqrt{(N-k)!}}\frac{(a^\dagger)^k}{\sqrt{k!}}\ket{0},
\end{equation}
where $a^\dagger$ and $b^\dagger$ are bosonic creation operators creating a particle in $\ket{2}$ and $\ket{1}$ respectively. The total angular momentum operators, defined as sums of the single atom operators $j_+ = \ket{2}\bra{1}$, $j_- = \ket{1}\bra{2}$ and $j_3 = \frac12(\ket{2}\bra{2} - \ket{1}\bra{1})$ running over the entire ensemble, can be expressed using the bosonic mode operators:
\begin{equation}
J_+ = a^\dagger b,\qquad J_- = ab^\dagger,\qquad J_3 = \frac12(a^\dagger a - b^\dagger b).
\end{equation}
This is referred to as the Schwinger representation. 

For large $N$ and states containing few $a$-excitations, i.e. states close to $\ket{1}^{\otimes N}$, the so-called Holstein-Primakoff approximation \cite{hp} can be derived by replacing $b$ and $b^\dagger$ with $\sqrt{N}$ and neglecting $\mathcal{O}(a^2)$-terms. The jump operators $J_+$ and $J_-$ are then proportional to $a^\dagger$ and $a$, respectively, and the transverse angular momentum components read
\begin{equation}
J_1 = \sqrt{\frac{N}{2}}X,\qquad J_2 = \sqrt{\frac{N}{2}}P ,
\end{equation}
where $X$ and $P$ denote ensemble quadratures defined analogously to CV phase space quadratures. 
Thus, the subset of the ensemble state space close to the fully polarized state behaves like a bosonic mode phase space, allowing to define CV states like the GKP basis states in the ensemble. 
We find that in order for this subset to be large enough to support sufficiently strongly squeezed GKP states, the number of atoms comprising the ensemble should be in the order of $10^3$ to $10^4$ (see App.~\ref{ensemblephysics}).

A quantum memory in the repeater context should also allow for performing operations like amplification and entanglement swapping that require interaction with an external electromagnetic field mode. As shown in Ref.~\cite{hammerer} and in App.~\ref{ensemblephysics}, suitable interactions between an ensemble and a target mode can be generated with the aid of an additional high intensity mediating field polarized along the quantization axis. 
Depending on the polarization of the target mode, the general expression
\begin{equation}
H_\text{eff} = \sum_{a,b \in \{1, 2\}}\omega_A^2\sum_j\frac{\mathbf{A}^{(-)}\cdot\braket{a|\mathbf{D}|e_j}\mathbf{A}^{(+)}\cdot\braket{e_j|\mathbf{D}|b}}{\Delta}\ket{a}\bra{b}
\end{equation}
simplifies to a Faraday-, a beam-splitter- or a two-mode squeezing interaction, as depicted in Table~\ref{tab::Atom:interactions}.

\begin{table*}
\begin{ruledtabular}
\begin{tabular}{c|cccc}
 Polarization & $y$ & $z$ & $\sigma^+$ & $\sigma^-$\\
\hline
Interaction & $pP$ & $pX$ & $a_La_A^\dagger + a_L^\dagger a_A$ & $a_La_A + a_L^\dagger a_A^\dagger$\\
\hline
Used for & Bell/syndrome measurement & Bell/syndrome measurement & Loss channel CC-amp. & Amplification channel
\end{tabular}
\end{ruledtabular}
\caption{\label{tab::Atom:interactions}Possible interactions between atomic ensemble and field mode in Holstein-Primakoff approximation. Quantization axis along $x$, circular polarization lying in $y$-$z$-plane. Upper case quadratures and subscript $A$ refer to the ensemble, lower case quadratures and subscript $L$ refer to the field mode.}
\end{table*}

\subsection{Entanglement swapping with atomic ensembles}
\label{subsec::ensoperations}
\begin{figure}
\includegraphics[width=0.45\textwidth]{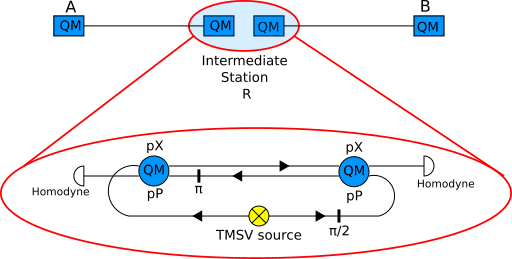}
\caption{TMSV-mediated entanglement swapping between ensemble modes. }
\label{fig::Rate:tmsvswapping}
\end{figure}

In Section~\ref{subsec::ESqubit} we showed that the information necessary for both entanglement swapping and QEC can be extracted by measuring the sum of $x$-quadratures and the difference of $p$-quadratures of the two center modes. We now demonstrate how this measurement can be implemented in the framework of atomic spin ensembles as memories using the Faraday interaction between the ensemble and modes of an optical two-mode squeezed vacuum (TMSV). The scheme is illustrated in Fig.~\ref{fig::Rate:tmsvswapping}. 

Since a standard technique for measuring quadratures of electromagnetic modes is available in the form of homodyne detection, the challenge lies in transferring the ensemble quadratures onto light modes in a suitable fashion. 
Though interactions like $pX$ or $pP$ allow to transfer one ensemble quadrature to the light mode, the remaining quadrature is simultaneously altered, as quadratures transform as 
\begin{eqnarray}
x_A^\prime =& x_A,\hphantom{ + x_B} \qquad &p_A^\prime = p_A - p_C, \nonumber\\
x_C^\prime =& x_A + x_C, \qquad &p_C^\prime = p_C
\end{eqnarray}
under $p_CX_A$ and
\begin{eqnarray}
x_A^\prime &= x_A + p_C, \qquad &p_A^\prime = p_A, \nonumber\\
x_C^\prime &= x_C + p_A, \qquad &p_C^\prime = p_C
\end{eqnarray}
under $p_CP_A$.
This becomes an issue when the second quadrature is read out onto a second light mode, since it still contains unwanted contributions from the first light mode. 
To avoid such effects, the light modes must be suitably correlated, e.g. as parts of a TMSV, where in the limit of infinite squeezing 
\begin{equation}
x_C - x_D = 0, \qquad p_C + p_D = 0 
\end{equation}
holds true for modes $C$ and $D$. 
We further assume that it is possible to rotate the vector of quadratures for light modes, i.e. performing transformations like $\begin{pmatrix}x\\p\end{pmatrix}\rightarrow\begin{pmatrix}p\\-x\end{pmatrix}$, e.g. using waveplates. \\

In the following we consider two modes $C$ and $D$ forming a TMSV, that pass through two atomic ensembles $A$ and $B$ in opposite directions. As we know, the interaction type depends on the fields' polarization. Here we set $C$ to be polarized in $y$-direction and $D$ to be polarized in $z$-direction. 
Initially, $C$ interacts with ensemble $A$ via $p_CX_A$ while simultaneously, $D$ rotated by $\frac{\pi}{2}$ interacts with $B$ via $p_DP_B$. After these interactions, the quadratures read
\begin{eqnarray}
q_A^\prime &= \begin{pmatrix}x_A\\p_A-p_C\end{pmatrix}, \quad &q_B^\prime = \begin{pmatrix}x_B-x_D\\p_B\end{pmatrix}, \nonumber\\
q_C^\prime &= \begin{pmatrix}x_C+x_A\\p_C\end{pmatrix},\quad &q_D^\prime = \begin{pmatrix}p_D+p_B\\-x_D\end{pmatrix}.
\end{eqnarray}

Next, $C$ passes through ensemble $B$ and in turn interacts via $p_CX_B$ , resulting in
\begin{equation}
q_B^{\prime\prime} = \begin{pmatrix}x_B-x_D\\p_B-p_C\end{pmatrix}, \quad q_C^{\prime\prime} = \begin{pmatrix}x_A+x_B+x_C-x_D\\p_C\end{pmatrix}.
\end{equation}
Due to the TMSV-correlations, the upper component of $C$ simplifies to $x_A + x_B$ and thus the aim of transferring the sum of $x$-quadratures to a light mode is achieved. 

Similarly, mode $D$ interacts with $A$ via $p_DP_A$ after undergoing another rotation by $\pi$, and one obtains
\begin{equation}
q_A^{\prime\prime} = \begin{pmatrix}x_A+x_D\\p_A-p_C\end{pmatrix}, \quad q_D^{\prime\prime} = \begin{pmatrix}p_A-p_B-p_C-p_D\\x_D\end{pmatrix},
\end{equation}
where the upper component simplifies to the desired $p_A - p_B$.
Both light modes now carry one of the necessary linear combinations of ensemble quadratures whose measurement can thus be performed by homodyning.

\section{Rate analysis}
\label{sec::rateanalysis}
In this section, we will examine and quantify the performance of a second-generation quantum repeater based on the principles outlined in the preceding section. The quantity of interest will be the secret key rate $S$ serving as a measure for the number of secure key bits one can generate using the repeater in a QKD scheme. Similar analyses have been carried out for memoryless repeaters based on optical GKP states \cite{gkp3rdgen1, alloptical, gkp3rdgen2, frank_quditgkprepeater}. Besides the seminal ``DLCZ'' quantum repeater proposal for atomic-ensemble memories \cite{dlcz} with built-in memory loss quantum error detection, but without any form of loss correction, there are numerous works on analyzing memory-based quantum repeaters with single-spin memories and no QEC on the stationary spin qubits. Here, for comparison, we shall consider the work of Ref.~\cite{rateanalysis} that uses single-spin memories without QEC. We refer to this scheme as the ``correctionless'' scheme.

\subsection{QKD secret key rates}
The secret key rate $S = rR$ is composed of the raw rate $R$ and the secret key fraction $r$. The former is given as the reciprocal of the time needed until a GKP Bell pair is distributed between Alice and Bob (or, more precisely, an effective GKP Bell pair when Alice and Bob measure their atoms immediately as they can do in a QKD scheme), whereas the latter depends on the specific QKD protocol and is given as $r = 1 - 2h(\text{QBER})$, with the binary entropy function
\begin{equation}
h(x) = -x\log_2(x) - (1-x)\log_2(1-x),
\end{equation}
for the case of BB84 and equal error probabilites for Pauli-$X$ and Pauli-$Z$. Here, QBER denotes the quantum bit error rate, i.e. the probability of the eventually and effectively distributed state being subject to a Pauli error. 
Of course the secret key fraction cannot reasonably take negative values, therefore it is set to 0 if $h(\text{QBER}) < \frac12$, resulting in a vanishing secret key rate $S$. It turns out that this is the case for a large portion of the domain, namely $0.11 < \text{QBER} < 0.89$, so that non-zero rates can only be achieved for $\text{QBER} \leq 0.11$.

To understand how the QBER arises we begin by going back to the entanglement distribution. As explained in Section~\ref{subsec::ESdis} it is a stochastic process following a geometric distribution with success probability $p = p_\text{link}e^{-L_0/L_\text{att}}$. As a consequence, successfully distributed segments may need to wait for neighboring segments to finish distribution, incurring losses from memory decoherence that are converted to Gaussian shifts via amplification. 
In Ref.~\cite{alloptical} it was shown that teleportation-based quantum error correction with a Bell pair whose modes are shifted by $\sigma_\text{Bell}^2$ is equivalent to correction with a perfect Bell pair where both the input- and output modes additionally undergo a shift of $\sigma_\text{Bell}^2$ before and after the correction process, respectively. 
In our case, the Bell pair is the more recently finished segment involved in the swapping process and $\sigma_\text{Bell}^2$ is the sum of the natural variance $ \delta^2$ and the variance corresponding either to the artificial loss channel in the case of CC-amplification or the memory loss incurred during one time step in the case of preamplification.
The total variance of the syndrome measured in the correction process is the sum of $\sigma_\text{Bell}^2$ and the variance carried by the input mode, i.e. $\sigma_\text{tot}^2 = 2\delta^2 + \mathds{E}(\sigma_\text{add}^2)$, where $\mathds{E}(\sigma_\text{add}^2)$ depends on the amplification strategy and is given according to the equations in Section~\ref{subsec::AS}.

The QBER quantifies the probability of the final state distributed between Alice and Bob containing an unheralded Pauli-$X$ error arising from misinterpretation of the syndrome information when performing error correction.  The choice of $X$ over $Z$ is arbitrary and has no impact on the result, as all processes such as Gaussian shifts and error correction are symmetrical in $X$ and $P$. 

A Pauli error can occur in each swapping process, namely if the true shift falls into the areas labeled ``error''  in Fig.~\ref{fig::GKP:fehler}. The probability can hence be found by integration:
\begin{equation}
p_\text{Pauli} = \sum_{k\in\mathds{Z}} \int_{(2k+1)\sqrt{\pi}-\sqrt{\pi}/2}^{(2k+1)\sqrt{\pi} + \sqrt{\pi}/2}dx\, \frac{1}{\sqrt{2\pi\sigma_\text{tot}^2}}\exp\left(-\frac{x^2}{2\sigma_\text{tot}^2}\right),
\label{eq::Rate:ppauli}
\end{equation}
which can be further simplified to
\begin{equation}
p_\text{Pauli} = 1 - \int_{-\sqrt{\pi}/2}^{\sqrt{\pi}/2}dx\, \frac{1}{\sqrt{2\pi\sigma_\text{tot}^2}}\exp\left(-\frac{x^2}{2\sigma_\text{tot}^2}\right),
\label{eq::Rate:ppauli2}
\end{equation}
as for the range of variances that allow for a non-zero secret key rate, the areas lying further out barely contribute to the sum.

Assuming $p_\text{Pauli}$ were identical for all swappings, the probability of a total of $j$ errors occurring, if possible correction steps on Alice and Bob's memories are disregarded as their atoms are measured immediately, would be given by the binomial distribution with parameters $n-1$ and $p_\text{Pauli}$. The final state contains an error if and only if the total number of errors is odd, since two Pauli errors cancel one another. Thus, the QBER reads
\begin{eqnarray}
\text{QBER} &= &\sum_{\substack{j = 0\\j \text{ odd}}}^{n-1}\begin{pmatrix}n-1\\j\end{pmatrix}p_\text{Pauli}^j(1-p_\text{Pauli})^{n-1-j}\nonumber\\
 & = &\frac12\left[1 - (1-2p_\text{Pauli})^{n-1}\right].
\label{eq::Rate:oddsum}
\end{eqnarray}  
Incidentally, it becomes apparent here why the $\text{QBER} \geq 0.89$ regime does not play a role:
Even in the limit  $\sigma_\text{tot} \rightarrow \infty$, corresponding to a uniform distribution of shifts, one would only have $p_\text{Pauli} = \frac12$. In that case, Eq.~\refer{eq::Rate:oddsum} simplifies to
\begin{equation}
\text{QBER} = \frac{1}{2^{n-1}}\sum_{\substack{j = 0\\j \text{ odd}}}^{n-1} \begin{pmatrix}n-1\\j\end{pmatrix} = \frac12,
\label{eq::Rate:norightside}
\end{equation}
yielding a QBER still far below the threshold of 0.89.

The appearance of a binomial distribution already imposes a bound on the achievable secret key rates, as the number of intermediate stations cannot be increased at will, at least in the case of finite GKP squeezing. 
This can be seen by considering the condition 
\begin{equation}
\frac12[1 - (1-2p_\text{Pauli})^{n-1}] \leq 0.11
\end{equation}
for a non-vanishing secret key rate, which can be rearranged for $p_\text{Pauli}$ to give an $n$-dependent threshold of how big the probability of a Pauli error during each swapping may be so that the rate does not vanish:
\begin{equation}
p_\text{Pauli} \leq \frac12\left(1 - 0.78^{1/(n-1)}\right).
\label{eq::Rate:pstrich}
\end{equation}
For $n \rightarrow \infty$ this curve approaches zero, whereas $p_\text{Pauli}$ exhibits a strictly positive lower bound due to the variance of finitely squeezed GKP states. Therefore there exists an $n$ where $p_\text{Pauli}$ ceases to lie below the curve from Eq.~\refer{eq::Rate:pstrich}, and for this and all higher $n$ it follows $S = 0$. 
In the regime below this maximal $n$, an increasing number of segments has positive as well as negative impacts on the secret key rate; in particular it improves $R$ but simultaneously raises the QBER. This suggests the existence of an optimal $n$, which will be confirmed in the further analysis below.

\subsection{Methods for repeater rate analysis}
\label{subsec::Methodik}
We consider five parameters that influence the secret key rate. Aside from the total distance $L$ and the segment number $n$, these are the link efficiency $p_\text{link}$, the variance of GKP states $\delta^2$ and the coherence time $t_\text{coh}$ of the quantum memories. 
One can try to construct an analytical expression for the secret key rate by suitably inserting the equations presented in the previous section into each other.
However, a difficulty arises from the treatment of random variables. The simplest approach is, of course, to replace random variables with their expectation value. 

For the raw rate
\begin{equation}
\overline{K_n} = \sum_{i=1}^n (-1)^{i+1}\begin{pmatrix}n\\i\end{pmatrix}\frac{1}{1-q^i }
\end{equation}  
is derived and employed in Refs.~\cite{nadja_rateanalysis, waitingtime_pra, rateanalysis} as the average number of time steps for the total process if the deterministic entanglement swapping is performed as soon as possible. The raw rate is then obtained from $R = 1/\overline{K_n}$. 
To find the secret key fraction we insert the average total variance into Eq.~\refer{eq::Rate:ppauli} as $\sigma_\text{tot}^2$ and subsequently calculate the QBER via the binomial distribution. The average variance, presented in more detail in Section~\ref{subsec::AS}, is found using the probability distribution for the waiting time, which can be understood as the absolute value of the difference of two geometrically distributed random variables. 

This ansatz exhibits two potential problems, however. Firstly, it assumes that the error probability is the same for each swapping, which is not necessarily the case, and secondly, the substitution of random variables by their expectation value is performed too early when the average additional variance is inserted into Eq.~\refer{eq::Rate:ppauli} instead of calculating the QBER separately for all possible variances and averaging afterwards. This is not mathematically correct, since the relation $\mathds{E}(f(X)) = f(\mathds{E}(X))$ only holds for linear $f$, and the QBER depends non-linearly on the variance. 

In order to estimate how grave these issues are, we additionally consider two numerical methods. As the first method, a simulation suggests itself due to the discretized nature of time in our considerations. At the beginning, a geometrically distributed random variable is drawn for each segment, governing how many time steps it will require to successfully distribute a Bell pair. Then in each time step, the possibility of a swapping is checked and the variance carried by the modes is increased where appropriate. When a swapping occurs, $p_\text{Pauli}$ is calculated and appended to a list that is evaluated at the end by drawing uniform random variables in $[0, 1]$ and comparing with the error probabilities. The parity of the total number of errors carries the information whether an error is contained in the final state, such that by averaging over 1000 iterations of this comparison, the QBER for the realization at hand of the $n$ random variables can be calculated. 
The total process is repeated 5000 times. 

However, this method is not entirely unproblematic in itself, since a number of 5000 repetitions is too low to cover all possible configurations of the $n$ geometric random variables. Even if instead of the concrete value only the relative order were relevant, this would result in $n!$ possibilities, which is more than 5000 for $n = 7$ already. An increase of the number of repetitions would further extend the already quite long runtime of the program though. 

The second method consists in averaging the QBER numerically using the probability distribution of the waiting time found in App.~\ref{subsec::appGeo}, where the theoretically infinite summation is truncated after 50000 summands. 
\begin{figure}
\includegraphics[width=0.45\textwidth]{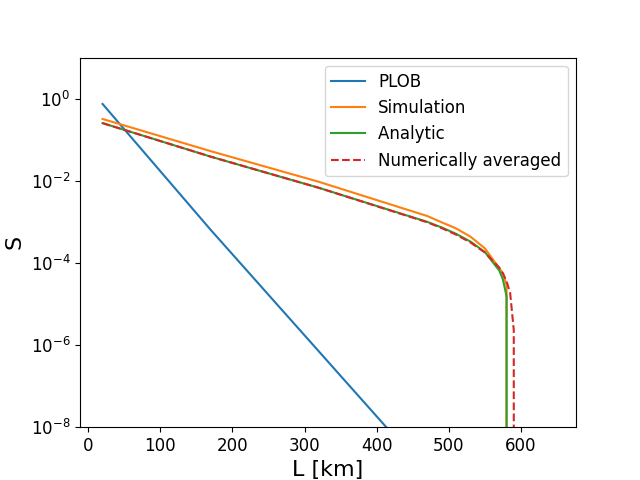}
\caption{Comparison of methods to determine secret key rate. All three approaches show fairly good agreement, with the simulation yielding slightly higher results at lower $L$ and the analytically averaged rate dropping off slightly later at higher $L$. Preamplification for each timestep is used as the amplification strategy for all methods. Parameters: $n=4$, $p_\text{link}=0.7$, $\delta^2=0.05$, $t_\text{coh}=10s$. }
\label{fig::Rate:4segs}
\end{figure}
Figure~\ref{fig::Rate:4segs} shows the rates calculated according to the analytical method as well as both numerical methods for a repeater with four segments. The comparisons shows that the analytical method yields almost the same results one obtains using numerical averaging, showing that the time of averaging plays a minor role. The simulation shows slightly higher rates than the other two methods, which might be ascribed to the restricted sample size, as comparatively unlikely configurations with long waiting times do not contribute.
Overall, no significant deviations have become manifest, however, so that the analytical approach is indeed suitable for the further rate analysis.

\subsection{Results}
\label{subsec::results}
Some first physical insights can also be drawn from Fig.~\ref{fig::Rate:4segs}. For instance it is apparent that at distances shorter than 50km the rates lie below the PLOB bound, meaning that the repeater is not advantageous in this case, whereas at longer distances the rates achieved with repeater surpass direct transmission by several orders of magnitude. 
Also interesting is a comparison with the results from Refs.~\cite{alloptical} and \cite{rateanalysis}. One observes that the second generation repeater yields considerably higher rates per time step than the memoryless repeater from Ref.~\cite{alloptical} when using no higher level encoding and a comparable number of intermediate stations, although this has to be adjusted for the different clock rates between purely optical and memory-based schemes.

\begin{table*}
\begin{ruledtabular}
\begin{tabular}{cc|cccccccc}
 & & \multicolumn{8}{c}{$n$}\\
 & & 2 & 4 & 8 & 16 & 32 & 64 & 128 & 256\\
\hline
\multirow{4}{*}{$\delta^2$} & 0.05 & 0.2075 & 0.0858 & 0.0390 & 0.0125 & $\leq$0.0010 & $\leq0.0010$ & $\leq0.0010$ & $\leq0.0010$\\
 & 0.03 & 0.2475 & 0.1258 & 0.0790 & 0.0525 & 0.0348 & 0.0220 & 0.0123 & 0.0046\\
 & 0.02 & 0.2675 & 0.1458 & 0.0990 & 0.0725 & 0.0548 & 0.0420 & 0.0323 & 0.0246\\
 & 0.01 & 0.2875 & 0.1658 & 0.1190 & 0.0925 & 0.0748 & 0.0620 & 0.0523 & 0.0446
\end{tabular}
\end{ruledtabular}
\caption{Thresholds for variance $\gamma^2$ introduced by noisy operations in the limit of infinite coherence time.\label{tab::Rate:gammathresh}}
\end{table*}

\begin{figure}
\includegraphics[width=0.49\textwidth]{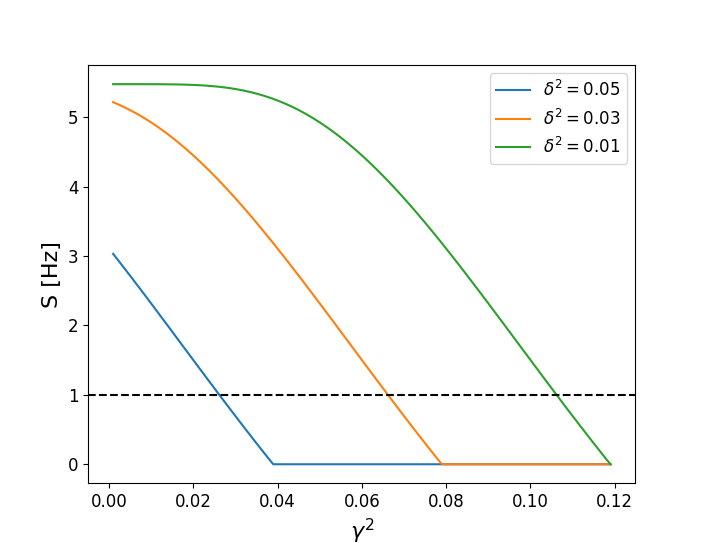}
\caption{Decrease of secret key rate at $L=800$km with 8 segments as a function of $\gamma^2$ under the assumption of perfect memories, i.e. $\alpha \rightarrow 0$. Other parameters: $p_\text{link} = 0.7$.}
\label{fig::Rate:gammatresh}
\end{figure}

When comparing with Ref.~\cite{rateanalysis} for the same values of $p_\text{link}$ and $t_\text{coh}$, the repeater using GKP error correction yields somewhat higher rates than the non-ideal case without correction; however, when considering the ideal case from Ref.~\cite{rateanalysis}, the results surpass our rates. In particular they drop off more slowly at increasing distances. This remains unchanged for an increased number of intermediate stations, since while our rate improves for fixed $L$ when increasing $n$, this effect occurs for the case without correction as well. The designations ideal and non-ideal refer to the parameter $\mu$ introduced in Ref.~\cite{rateanalysis}, which quantifies possible depolarization arising from the application of gates as well as imperfect initial states. Effects of this kind can be modeled in our scheme by introducing an additional variance $\gamma^2$ in each swapping process. In analogy to the table of $\mu$ thresholds in Ref.~\cite{rateanalysis}, Table~\ref{tab::Rate:gammathresh} shows the maximal values that $\gamma^2$ can take such that a non-vanishing rate can still be achieved, for the limit of infinite coherence time, i.e. $\alpha \rightarrow 0$ and $\sigma^2_\text{tot} = 2\delta^2 + \gamma^2$.
The decrease of the secret key rate with increasing $\gamma^2$ is visualized in Fig.~\ref{fig::Rate:gammatresh} for different GKP squeezing values.

\begin{figure*}
\begin{minipage}{0.45\textwidth}
\includegraphics[width=\textwidth]{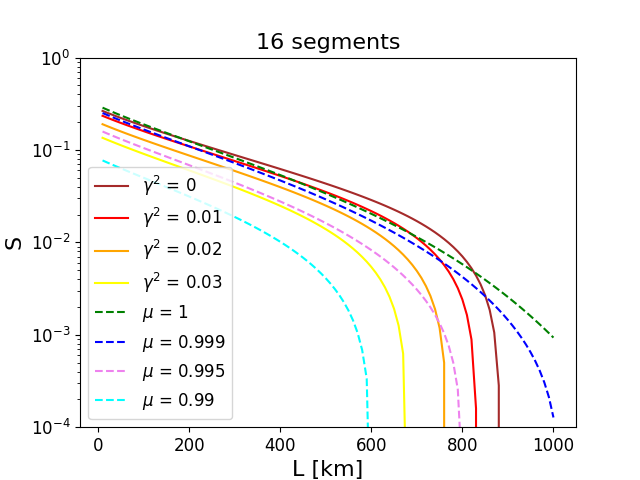}
\end{minipage}
\begin{minipage}{0.45\textwidth}
\includegraphics[width=\textwidth]{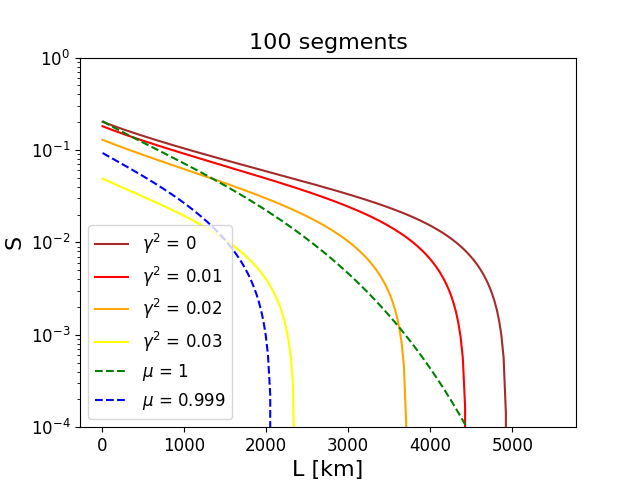}
\end{minipage}
\caption{Comparison of GKP-based protocol with Kamin et al. \cite{rateanalysis} for non-ideal noise parameters $\gamma^2$ and $\mu$. Other parameters: left: $\delta^2 = 0.03$, $t_\text{coh} = 0.1$s, $p_\text{link} = 0.7$; right: $\delta^2 = 0.02$, $t_\text{coh} = 0.1$s, $p_\text{link} = 0.7$.    }
\label{fig::Rate:finitegammamu}
\end{figure*}

Additionally, the rates for some non-ideal values of $\gamma^2$ and $\mu$ are shown in Fig.~\ref{fig::Rate:finitegammamu}. For a 16-segment repeater with a coherence time of 0.1s and a GKP squeezing of $\delta^2=0.03$ one observes approximately equal rates for $\mu = 0.995$ and $\gamma^2 = 0.03$ as well as $\mu = 0.999$ and $\gamma^2 = 0.01$ when $L$ is not too high. As $L$ is increased, the correctionless rates drop off slower than the corresponding GKP-corrected rates, with the result that the maximal distance achievable with $\mu=0.995$ lies between the values for $\gamma^2 = 0.02$ and $\gamma^2 = 0.01$, and $\mu = 0.999$ surpasses even the idealized case of $\gamma^2$ in this respect. For a repeater with 100 segments and a higher GKP squeezing of $\delta^2 = 0.02$ the rate with $\gamma^2 = 0.02$ surpasses $\mu = 0.999$ at all total lengths, and the maximal distance with $\mu = 0.999$ can be surpassed even with $\gamma^2 = 0.03$. A fundamental difference between the schemes is that in the GKP case higher noise parametrized by $\gamma^2$ can be offset by a better squeezing $\delta^2$, whereas in the scheme from Ref.~\cite{rateanalysis} no such possibility exists. The GKP scheme thus appears to exhibit a better resilience to noisy operations, with the caveat that $\mu$ and $\gamma^2$ correspond to different physical processes and it is not immediately obvious how experimentally challenging it is to achieve a certain numerical value in each case. 
\begin{figure}
\includegraphics[width=0.45\textwidth]{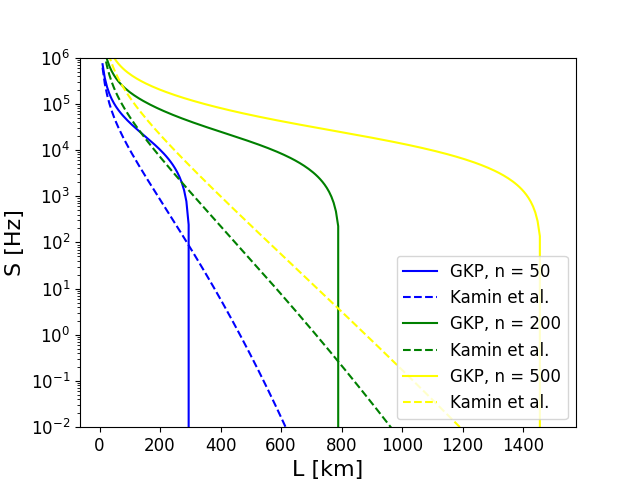}
\caption{Rates with and without error correction at a comparatively short coherence time of 1ms. As more intermediate stations are introduced, the GKP scheme gains an advantage over the correctionless scheme.  Parameters: $\delta^2 = 0.02$, $p_\text{link} = 1$, $t_\text{coh} = 0.001$s.}
\label{fig::Rate:lowcoherence}
\end{figure}

\begin{figure}
\includegraphics[width=0.45\textwidth]{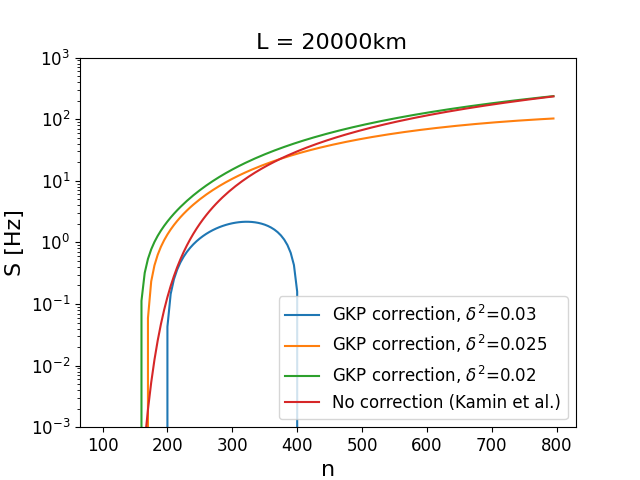}
\caption{Modified secret key rate as a function of $n$ for L=20000 km, corresponding to the longest reasonable distance on earth. Error correction enables higher rates compared to the correctionless scheme when using few intermediate stations, whereas at higher $n$ the two schemes barely differ. Parameters: $p_\text{link} = 0.7$, $t_\text{coh} = 10$s. }
\label{fig::Rate:20000}
\end{figure}
However, in the remainder of this work we set $\gamma^2 = 0$, meaning that comparability is only given for the case $\mu = 1$. Thus using GKP error correction does not improve the secret key rate at least for the parameter values considered in Fig.~\ref{fig::Rate:4segs}.

Despite the four-segment chain with low squeezing not being able to surpass the correctionless scheme, regimes can be found where GKP-correction can lead to significantly higher rates, as illustrated in Fig.~\ref{fig::Rate:lowcoherence}. There we consider a comparatively short coherence time of 1ms and high number of segments and observe that rate improvements of several orders of magnitude can be achieved at some total distances. The advantage of error correction becomes more pronounced as the number of segments is increased both in terms of key rate ratio to the correctionless case as well as range of distances at which an improvement is possible. For a fixed number of segments, the ratio is highest when each segment is as long as it can possibly be while maintaining a non-zero rate. 

Incidentally, exact expressions are available in Ref.~\cite{rateanalysis} only for 4 or fewer segments, since the probability generating functions for the total waiting time become extremely complicated for higher $n$ and without any additional simplifying assumptions, resulting in no material being available for comparison with 16 or more segments. However, when considering the total waiting time a sum of $n-1$ independent random variables, each of which is the absolute value of the difference of two geometric random variables, an approximation for the expectation value $\mathds{E}(e^{-\alpha D_n})$, with $D_n$ denoting the waiting times summed over all segments, can be constructed:
\begin{equation}
\mathds{E}(e^{-\alpha D_n}) = \left(\frac{p^2}{1-q^2}\right)^{n-1} \left(\frac{1 + qe^{-\alpha}}{1 - qe^{-\alpha}}\right)^{n-1}.
\label{eq::Rate:kaminana}
\end{equation} 
The derivation of this expression is presented in  App.~\ref{waitingstatistics} and relies on the simplifying assumptions \cite{rateanalysis, Goodenough2024} of (i) deterministic entanglement swapping, (ii) swapping as soon as possible, and (iii) Alice and Bob measuring their atoms immediately in a QKD application, thus freeing them of any storage-induced loss errors occurring at the repeater's sender and receiver stations. 
Note that the definition of $D_n$ used in our derivation deviates slightly from the one used in Ref.~\cite{rateanalysis}; for more details see App.~\ref{waitingstatistics}.
Additionally, Eq.~\refer{eq::Rate:kaminana} is based on the assumption that the $n-1$ random variables constructed as the absolute values of the differences of two geometrically distributed random variables are statistically independent, which is not actually the case. However, comparison with numerical simulations shows that this has a negligible impact on the resulting expectation value, at least for the relevant values of $\alpha$ and $q$, such that Eq.~\refer{eq::Rate:kaminana} constitutes a valid approximation.
Using this expectation value, the QBER and hence the secret key rate for the correctionless case can be calculated as explained in Ref.~\cite{rateanalysis}. 

\begin{figure*}
\begin{minipage}{0.45\textwidth}
\includegraphics[width=\textwidth]{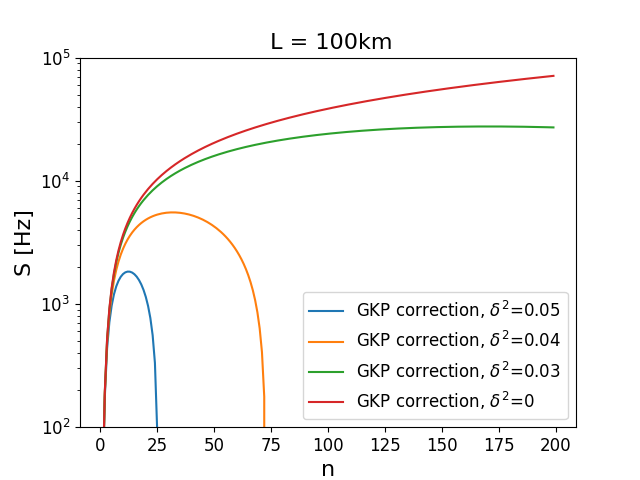}
\end{minipage}
\begin{minipage}{0.45\textwidth}
\includegraphics[width=\textwidth]{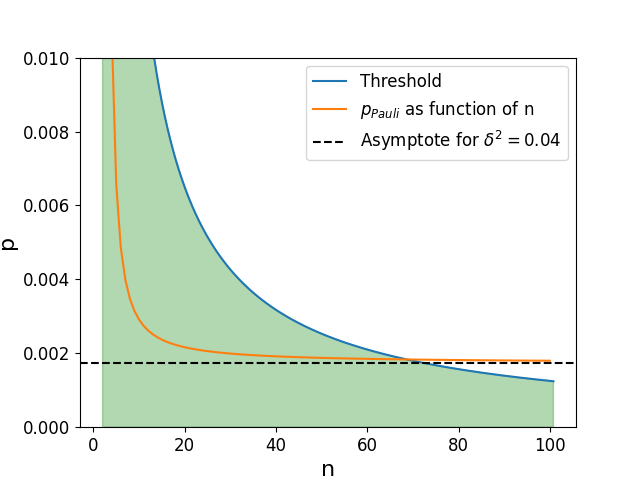}
\end{minipage}
\caption{Left: Influence of GKP squeezing on the modified secret key rate. Parameters: $L = 100$km, $p_\text{link} = 0.7$, $t_\text{coh} = 10$s. 
Right: Cause of the strong influence: The blue curve corresponds to Eq.~\refer{eq::Rate:pstrich}, the orange curve shows the true course of $p_\text{Pauli}$ as a function of $n$, with the dotted line as an asymptote in the limit of large $n$. If $\delta^2$ is varied, the asymptote shifts vertically and therefore changes the $n$-coordinate of the intersection of the blue and orange curves.  }
\label{fig::Rate:deltas100}
\end{figure*}

We are now also in a position to check the assumption of the existence of an optimal number of segments expressed above. In order to compare rates at different $n$ it is helpful to slightly modify the definition of the secret key rate to account for the dependence of the duration of a time step on $n$. We define the modified rate as 
\begin{equation}
S[\text{Hz}] = \frac{S}{\tau} = \frac{Scn}{L},
\end{equation}
which now carries the unit bits per second, or Hertz. 
For fixed total length $L$ one can plot the rates for different $n$ and search for a maximum, as illustrated in Fig.~\ref{fig::Rate:deltas100}. It turns out that as expected, a maximum can be found for all distances.  
Up to a certain length of about 50 km, the maximal rate is obtained with just two stations, corresponding to direct transmission between Alice and Bob.
For higher distances, intermediate stations improve the rates and the ideal number grows in a sublinear fashion.

Another interesting aspect is the influence of the GKP squeezing on the performance, as it constitutes the only parameter not appearing in Ref.~\cite{rateanalysis} and can therefore be at least theoretically chosen at will while other parameters are fixed for a comparison. 
Indeed it turns out that an improved squeezing can result in a vastly increased secret key rate, as apparent in Figs.~\ref{fig::Rate:20000} or \ref{fig::Rate:deltas100}. 

\begin{figure*}
\begin{minipage}{0.54\textwidth}
\includegraphics[width=\textwidth]{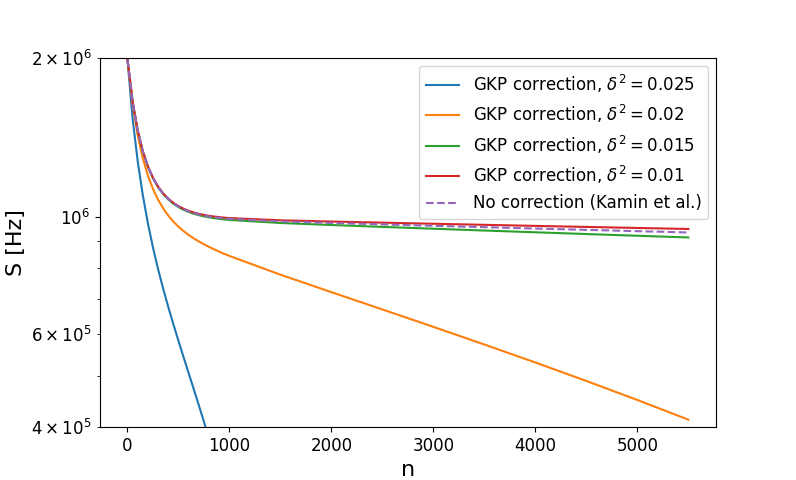}
\end{minipage}
\begin{minipage}{0.44\textwidth}
\includegraphics[width=\textwidth]{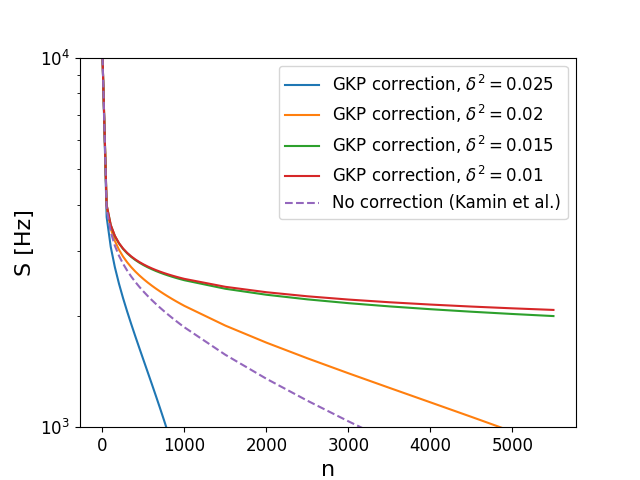}
\end{minipage}
\caption{Rates at constant $L_0$ (left: $L_0 = 0.1$km, right: $L_0 = 10$km) with $p_\text{link} = 1$, $t_\text{coh} = 1s$.}
\label{fig::Rate:l0feld}
\end{figure*}

\begin{figure*}
\begin{minipage}{0.45\textwidth}
\includegraphics[width=\textwidth]{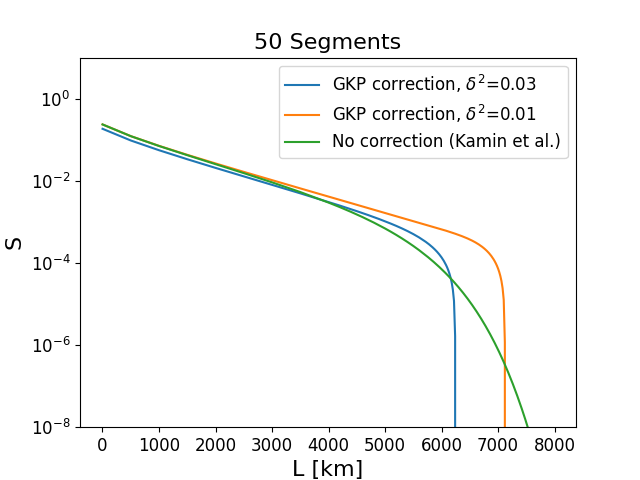}
\end{minipage}
\begin{minipage}{0.45\textwidth}
\includegraphics[width=\textwidth]{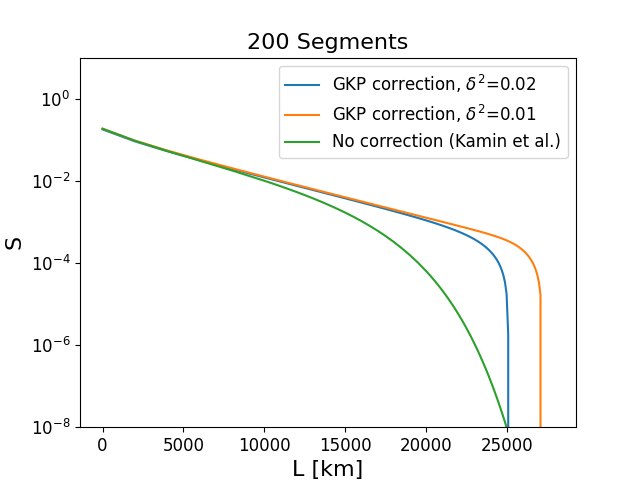}
\end{minipage}
\caption{Secret key rates with 50 and 200 segments respectively. Using strongly squeezed GKP states allows to achieve better rates at longer distances, as the secret key fraction stays close to 1 longer compared to the scheme without error correction. Other parameters: $p_\text{link} = 0.7$, $t_\text{coh} = 10$s.}
\label{fig::Rate:50200}
\end{figure*}

There the rates for $L$ fixed at 100km and various values of $\delta^2$ are plotted against $n$. One observes the maximum rate shifting position to higher $n$ and simultaneously reaching higher values if the squeezing is improved; additionally, the rate drops off more slowly after having passed the optimal $n$.

This strong dependence is caused by the fact that $\delta^2$ constitutes the lower bound for the variance of the Gaussian whose integration gives rise to the Pauli error probability. Therefore, no matter how short each segment is chosen, $p_\text{Pauli}$ can never fall below the value associated with $\delta^2$. However, as discussed in Section~\ref{subsec::Methodik}, the secret key fraction vanishes when the threshold curve derived from the binomial distribution is crossed in the $p$-$n$-diagram. The lower the bound created by $\delta^2$, the farther to the right this intersection will lie, as illustrated on the right-hand side in Fig.~\ref{fig::Rate:deltas100}. 
Better squeezing hence allows more segments to be used without detriment to the secret key fraction. In the limit of infinite squeezing the rate never reaches a maximum but keeps rising as more stations are added. 

It is also instructive to observe the rates achieved with different squeezing at a fixed segment length $L_0$. Fig.~\ref{fig::Rate:l0feld} shows the results for $L_0 = 0.1$km as well as $L_0 = 10$km, now using $p_\text{link} = 1$, in order to retain comparability with Ref.~\cite{rateanalysis}. 
The first thing to notice is that none of the curves is constant, reinforcing the fact that the rate does not just depend on $L_0$, but also in a non-trivial way depends on $n$. All curves in particular show a fast drop from the initial maximal value caused by the raw rate. 
The curves for comparatively low squeezing drop relatively fast with increasing $n$, in accordance with the discussion above, whereas for a squeezing of ca. 15dB or higher an approximately constant rate is obtained after the initial drop-off. The value approached by both the correctionless rate and the rate with highly squeezed GKP error correction lies at about half the classical limit $c/L_0$ for $L_0 = 0.1$km, and at about 10\% for $L_0 = 10$km. Thus at least with very favorably chosen parameters, rates in the order of magnitude of classical communication can in principle be achieved. 
Lastly note that the correctionless case can be surpassed only at high $L_0$ using very strong squeezing. 

For a suitable choice of $L$ and $n$ as well as high squeezing it is even possible to find rates that are several orders of magnitude higher than those obtained without error correction, as demonstrated in Fig.~\ref{fig::Rate:50200}. This vast improvement is caused by the different curve shapes of the secret key fraction $r$ when plotted as a function of $L$. Error correction allows $r$ to stay close to 1 as $L$ increases before dropping off very abruptly, in contrast to the behavior without correction, where the curve starts dropping earlier albeit more slowly. Close to the falling edge the rates with correction hence lie significantly higher. 

This effect extends so far as to make 200 segments sufficient to keep the secret key fraction close to 1 at a squeezing of 0.02, or 14dB, when bridging the longest reasonable distance imaginable on earth, namely the point-to-point connection of two antipodal points on the equator, separated by approximately 20000km.  
However, the resulting rate in this case still only lies at ca. 10Hz due to the low raw rate and the correction factor accounting for the duration of a time step scaling as $1/L_0$. 

Adding more intermediate stations would improve the rate, but reduce the advantage over the scheme from Ref.~\cite{rateanalysis} not requiring QEC, since this advantage is restricted to operation with few intermediate stations at a long total distance. In practice, though such an operation would reduce cost, it would also impose restrictions on the achievable rates. Thus a weighing between low costs and high transmission rates must be performed, and error correction following the scheme presented here is only useful when choosing the former.  \\

\section{Conclusions}
A second-generation quantum repeater with error correction based on the GKP code is theoretically possible and can be physically implemented using atomic spin ensembles as quantum memories. A subset of the ensemble state space can be described as a CV phase space, allowing to define GKP states and modeling noise as a bosonic loss channel. With the help of a high intensity mediating field, interactions of the Faraday-, beam-splitter and two-mode squeezing type arise naturally between the ensemble and a quantum light mode. The Faraday interaction, together with an optical ancilla TMSV, can be used to extract the necessary linear combinations of ensemble quadratures to perform both the syndrome measurement and the entanglement swapping. The two-mode squeezing and beam-splitter interactions enable amplification. 

The achievable secret key rates are strongly dependent on the GKP squeezing: in all examined cases, the rates at $\delta^2 = 0.05$ or 10dB GKP squeezing were lower than the corresponding rates of an existing correctionless scheme based on single-spin quantum memories, whereas with a squeezing of at least $\delta^2 = 0.02$ or 14dB significantly better results can be achieved using GKP error correction, especially at high segment length $L_0$. Note that these squeezing values only account for the GKP squeezing; however, technically the finite squeezing of the optical TMSV also causes contributions to $\delta^2$, meaning that the individual squeezing values of GKP and TMSV need to be higher than $\delta^2$. Assuming for simplicity that both contribute equally to $\delta^2$, their true squeezing would have to be $\delta^2/2$, or in other words, their squeezing parameter would have to be increased by 3dB compared to the nominal values found throughout this work.
At finite squeezing, the rate achievable at a fixed total distance is bounded from above, since the number of intermediate stations cannot be increased at will. Instead, the rate reaches a maximum at some segment number $n$ and starts decreasing as more stations are added. 
In the limit of infinite squeezing, this constraint is not present. 
An interesting topic for further research might be whether the demands on the squeezing can be reduced using techniques like the Highly Reliable Measurement \cite{alloptical} or a higher level QEC code on top of the GKP code. \\

\bibliography{literatur}
\bibliographystyle{unsrt}

\begin{acknowledgments}

We acknowledge funding from the BMBF in Germany
(QR.X, QuKuK) and from the Deutsche Forschungsgemeinschaft (DFG,
German Research Foundation) -- Project-ID 429529648 -- TRR 306 QuCoLiMa
("Quantum Cooperativity of Light and Matter'').
We further acknowledge support from the European Union and the
BMBF through QuantERA (ShoQC), and from the European Union’s
HORIZON Research and Innovation Actions (CLUSTEC).

\end{acknowledgments}

\appendix
\section{Atomic ensembles as quantum memories}
\label{ensemblephysics}

\subsection{Formalism}
The electromagnetic field in the absence of charges can be fully described by the vector potential $\mathbf{A}$, whose dynamics are governed by Maxwell's equations. 
In Coulomb gauge $\bm{\nabla}\cdot \mathbf{A} = 0$ the equation of motion takes the form of a wave equation. Although its solutions are in general a superposition of many frequency components, we will restrict ourselves here to the special case of a single frequency $\omega$. The vector potential can then be written \cite{oregonscript} as $\mathbf{A} = \tilde{\alpha} e^{-i\omega t}\mathbf{f}(\mathbf{r}) + \tilde{\alpha}^\ast e^{i\omega t}\mathbf{f}^\ast(\mathbf{r})$, where $\tilde{\alpha}$ is the amplitude and $\mathbf{f}(\mathbf{r})$ is a vector valued mode function containing the spatial dependency and satisfying $\int d^3r\, |\mathbf{f}(\mathbf{r})|^2 = 1$ and  $(\bm{\nabla}^2 + (\omega/c)^2)\mathbf{f}(\mathbf{r}) = 0$ .    
The electromagnetic fields of relevance to us would in practice be created by lasers and can be described as polarized plane waves, if details like the exact form of the Gaussian beam profile are ignored. Mathematically this means that the mode function takes the simple form  $\mathbf{f}(\mathbf{r}) \propto \bm{\epsilon} e^{i\mathbf{k}\cdot \mathbf{r}}$, with the polarization vector $\bm{\epsilon}$ and the wave vector $\mathbf{k}$. 

Quantization of the field is achieved by replacing the rescaled amplitudes $\alpha = \tilde{\alpha}\sqrt{2\omega V\epsilon_0}$ and $\alpha^\ast = \tilde{\alpha}^\ast\sqrt{2\omega V\epsilon_0}$ with bosonic creation and annihilation operators $a^\dagger$ and $a$. The quantized vector potential is then given by
\begin{equation}
\mathbf{A}(\mathbf{r}) = \sqrt{\frac{1}{2\omega V\epsilon_0}}(\bm{\epsilon} e^{i\mathbf{k}\cdot \mathbf{r}}a + \bm{\epsilon}^\ast e^{-i\mathbf{k}\cdot \mathbf{r}}a^\dagger),
\end{equation}
with the vacuum permittivity $\epsilon_0$ and the quantization volume $V$. In the following, we will abbreviate the prefactor by $\sqrt{\frac{1}{2\omega V\epsilon_0}} = C$ and write $\mathbf{A}$ as $\mathbf{A} = C(\bm{\epsilon} a + \bm{\epsilon}^\ast a^\dagger)$ without spatial dependency. For the case of interaction with a single atom, this simplification, often termed the dipole approximation, is unproblematic, provided that the light's wavelength is much larger than the typical atom radius. 
It is common to denote the summands constituting the quantized vector potential as $\mathbf{A}^{(+)} = C\bm{\epsilon} a$ and $\mathbf{A}^{(-)} = C\bm{\epsilon}^\ast a^\dagger$.\\

Next we turn to the description of the atoms: we consider a single atom as a two-level system with the two orthogonal states $\ket{1}$ and $\ket{2}$. The resulting operators $j_+ = \ket{2}\bra{1}$, $j_- = \ket{1}\bra{2}$ and $j_3 = \frac{1}{2}(\ket{2}\bra{2} - \ket{1}\bra{1})$ fulfill the commutation relations
\begin{equation}
[j_+, j_-] = 2j_3, \qquad [j_3, j_+] = j_+, \qquad [j_3, j_-] = -j_-, 
\end{equation}
known from angular momentum operators. In the case that $\ket{1}$ and $\ket{2}$ are magnetic sublevels to the same total angular momentum, $j_3$ can indeed be understood as the angular momentum component along the quantization axis. Correspondingly, the other angular momentum components can be defined as  $j_1 = \frac{1}{2}(j_+ + j_-)$ and $j_2 = \frac{1}{2i}(j_+ - j_-)$.
If $N$ such atoms are combined, the total angular momentum reads $\mathbf{J} = \sum_{i=1}^N \mathbf{j}_i$, or in components $J_+ = \sum_{i=1}^N j_{+, i}$, $J_- = \sum_{i=1}^N j_{-, i}$ and $J_3 = \sum_{i=1}^N j_{3, i}$, where the index $i$ runs over the atoms in the ensemble. These ensemble operators again fulfill the angular momentum algebra
\begin{equation}
\label{eq::Atom:drehalg}
[J_+, J_-] = 2J_3, \qquad [J_3, J_+] = J_+, \qquad [J_3, J_-] = -J_-.
\end{equation}
The state space of the ensemble is the $N$-fold tensor product of the two-dimensional one-particle Hilbert space; however, pure product states of the basis vectors $\ket{1}_i$ and $\ket{2}_i$ are not necessarily eigenstates of the total angular momentum. Instead the states space decomposes into orthogonal subspaces associated to the angular momentum values  $N/2, N/2 - 1, N/2 - 2, ...$, whose bases are given by complicated linear combinations of the product states. In the following we will consider the subspace associated to angular momentum $N/2$. It is spanned by completely symmetrized product states
\begin{equation}
\ket{\sigma_1, ..., \sigma_N} = \frac{1}{N!}\sqrt{\binom{N}{k}}\sum_{\pi\in S_N}\ket{\sigma_{\pi(1)}}\otimes...\otimes\ket{\sigma_{\pi(N)}},
\label{eq::Atom:symm}
\end{equation} 
where  $\sigma_i \in \{1, 2\}$, $k$ is the number of atoms in state $\ket{2}$ and $S_N$ denotes the symmetric group of $N$ elements, i.e. the set of permutations over the collection of atoms. Because of the symmetry and the fixed particle number $N$, such a state is uniquely specified by $k$, for which reason we can write $\ket{k}$ as a shorthand for Eq.~\refer{eq::Atom:symm}. 

Formally these states resemble those of a bosonic many-particle system and can thus be described using bosonic creation operators \cite{lightmed}:
\begin{equation}
\ket{k} = \frac{(b^\dagger)^{N-k}}{\sqrt{(N-k)!}}\frac{(a^\dagger)^k}{\sqrt{k!}}\ket{0}.
\end{equation}
In this representation, $a^\dagger$ creates a particle in state $\ket{2}$ while $b^\dagger$ creates a particle in state $\ket{1}$. As bosonic mode operators, $a$ and $b$ obey the usual commutation relations, and additionally fulfill $[a, b] = [a^\dagger, b^\dagger] = 0$.
 The angular momentum operators can also be represented using bosonic mode operators. In this so called Schwinger representation, the ladder operators take the form
\begin{equation}
J_+ = a^\dagger b, \qquad
J_- = ab^\dagger
\end{equation}
and raise or lower the angular momentum projection along the quantization axis by one unit respectively, destroying a $\ket{1}$ particle and creating a $\ket{2}$ particle or vice versa. The 3-component of angular momentum corresponds to half the difference between occupation numbers of the levels:
\begin{equation}
J_3 = \frac{1}{2}(a^\dagger a - b^\dagger b).
\end{equation}
It is easy to check that the angular momentum algebra is still fulfilled in Schwinger representation.  \\

For large $N$, states with few $a$-excitations can be simplified by approximating the operator $b^\dagger b$ with $N$, as well as $b$ and $b^\dagger$ with $\sqrt{N}$, while terms of squared order in $a$ are neglected. This is referred to as the Holstein-Primakoff approximation \cite{hp} and results in the form
\begin{equation}
J_+ = \sqrt{N}a^\dagger, \qquad J_- = \sqrt{N}a, \qquad J_3 = -\frac{N}{2}
\end{equation}  
for the angular momentum operators. The 1- and 2-components are therefore proportional to the quadratures of mode $a$:
\begin{equation}
\label{eq::Atom:HPquad}
J_1 = \sqrt{\frac{N}{2}}X, \qquad J_2 = \sqrt{\frac{N}{2}}P.
\end{equation}
The size of the corresponding bosonic phase space can be estimated via a geometric consideration:
the vector of total angular momentum lies on a sphere whose radius $\sqrt{(\frac N2 + 1)\frac N2}$ can be approximated as $N/2$ for large $N$. The Holstein-Primakoff approximation is valid for states that do not deviate too far from the fully polarized state $\ket{0} = \frac{(b^\dagger)^N}{\sqrt{N!}}$, i.e. for points on the sphere with a small polar angle $\theta$, where the sphere's negligible curvature allows to approximate it as a plane. As a bound for the validity of this approximation one can use $\theta_\text{max} = $10° as usual for small angle approximations. 
The transversal angular momentum components can be found using trigonometry. Using $\sin\theta = \theta$ one finds 
\begin{equation*}
J_\text{1, 2 max} = \theta_\text{max}\frac{N}{2}
\end{equation*}
for the maximum values within the region of validity.
Combined with Eq.~\refer{eq::Atom:HPquad} this implies
\begin{equation}
r_\text{HP} = \theta_\text{max}\sqrt{\frac N2}
\end{equation}
for the radius of the region of validity and thus the size of the available phase space. 

This is of interest in our repeater protocol as the variance $\Delta^2/4$ of the enveloping Gaussian of the Wigner function of GKP states must not be chosen too large, in order to avoid going beyond the region of validity. As a condition one could require that at $r_\text{HP}$ the Gaussian have dropped off by at least a factor of $1/e$ compared to the maximal value, leading to the relation $\Delta^2/4 \leq r_\text{HP}^2/2$. Because of $\delta^2 = 1/\Delta^2$ this implies a lower bound for the single peak variance:
\begin{equation}
\delta^2 \geq \frac{1}{N\theta_\text{max}^2}.
\end{equation}

As we saw in Section~\ref{subsec::results}, variances lower than 0.03 are desirable for a reasonable operation of the repeater, which means that the number of atoms should be in the order of magnitude of $10^3$ to $10^4$.

\subsection{Light-matter interaction}
We start by considering a single atom interacting with a quantized light mode of frequency $\omega_L$. The total system's Hamiltonian can be decomposed as $H = H_{\text{atom and int}} + H_{\text{field}}$, where the atomic and interaction part
\begin{equation}
H_{\text{atom and int}} = \frac{(\mathbf{p} + e\mathbf{A})^2}{2m_e} + V(\mathbf{x})
\end{equation}
arises from the free atomic Hamiltonian by minimal coupling.
Here we assumed that only one electron contributes significantly to the interaction, as is the case e.g. in alkali metals, and that the light's wavelength is much larger than the atom radius, allowing to drop the spatial dependency of $\mathbf{A}$. 
Expanding the binomial yields three terms: Firstly, one recovers the free atomic Hamiltonian from the potential and the $\mathbf{p}^2$ contribution, secondly, one finds a term  $\frac{e}{m_e} \mathbf{p}\cdot \mathbf{A}$ that can be interpreted as an interaction Hamiltonian, and thirdly the term $\frac{e^2}{2m_e}\mathbf{A}^2$ appears, which does not contain any degrees of freedom from the atom and can typically be neglected if the light's intensity is sufficiently weak. 
Thus, in total one finds 
\begin{equation}
H = \underbrace{\frac{\mathbf{p}^2}{2m_e} + V(\mathbf{x})}_{H_\text{atom}} + \underbrace{\frac{e}{m_e} \mathbf{p}\cdot \mathbf{A}}_{H_\text{int}} + H_{\text{field}}
\end{equation}
for the Hamiltonian.\\

In order to calculate the interaction, it is helpful to rewrite $\mathbf{p}$ as $\mathbf{p} = im_e[H_\text{atom}, \mathbf{x}]$. $H_\text{atom}$ can also be expressed in spectral decomposition as
\begin{equation*}
H_\text{Atom}= \sum_i \omega_i \ket{\phi_i}\bra{\phi_i},
\end{equation*}
where $\omega_i$ is the energy of the $i$-th eigenstate $\ket{\phi_i}$ and the sum runs over all relevant energy levels of the atom. Here, relevant means that there exists another state such that the energy difference does not deviate too much from the light mode's frequency. Typically this condition will lead to level schemes consisting of two degenerate energy levels containing several magnetic sublevels. In the following, states of the lower level will be denoted as $\ket{i}$, those of the higher level as $\ket{e_i}$ and their energy difference as $\omega_A$.
Inserting the spectral decomposition we find for $\mathbf{p}$:
\begin{widetext}
\begin{eqnarray}
\mathbf{p} &= &im_e\left[ \sum_i \omega_i \ket{\phi_i}\bra{\phi_i}, \mathbf{x}\right] =  im_e\left[ \sum_i \omega_i \ket{\phi_i}\bra{\phi_i}, \sum_{jk}\braket{\phi_j|\mathbf{x}|\phi_k}\ket{\phi_j}\bra{\phi_k}\right]\nonumber \\
&= &im_e\sum_{ijk}\braket{\phi_j|\mathbf{x}|\phi_k}\omega_i(\delta_{ij}\ket{\phi_i}\bra{\phi_k} - \delta_{ik}\ket{\phi_j}\bra{\phi_i})\nonumber\\
 &= &im_e\sum_{ij}\omega_i(\braket{\phi_i|\mathbf{x}|\phi_j}\ket{\phi_i}\bra{\phi_j} - \braket{\phi_j|\mathbf{x}|\phi_i}\ket{\phi_j}\bra{\phi_i}),
\end{eqnarray} 
\end{widetext}
where one can now introduce the total dipole operator $\mathbf{D} = -e\mathbf{x}$ and partial operators $\mathbf{D}_{ij} = \mathbf{D}_{ij}^{(-)} + \mathbf{D}_{ij}^{(+)} = \braket{e_j|\mathbf{D}|i}\ket{e_j}\bra{i} + \braket{i|\mathbf{D}|e_j}\ket{i}\bra{e_j}$ for each transition  $\ket{i} \leftrightarrow \ket{e_j}$ to obtain
\begin{equation}
H_\text{int} = -i\mathbf{A}\omega_A\sum_{ij}(\mathbf{D}_{ij}^{(-)} - \mathbf{D}_{ij}^{(+)})
\end{equation}
for the interaction Hamiltonian, since the dipole operator's matrix elements are non-zero only for transitions between the upper and the lower level. 

In the interaction picture, this can be further modified by performing a rotating-wave approximation: the light mode- and dipole operators transform as $a(t) = a(0)e^{-i\omega_L t}$ and $a^\dagger(t) = a^\dagger(0)e^{i\omega_L t}$, as well as $\mathbf{D}_{ij}^{(-)}(t) = \mathbf{D}_{ij}^{(-)} (0)e^{i\omega_At}$ and $\mathbf{D}_{ij}^{(+)}(t) = \mathbf{D}_{ij}^{(+)}(0)e^{-i\omega_At}$ respectively, so that terms like  $\mathbf{A}^{(+)}\cdot\mathbf{D}_{ij}^{(+)}$ that do not conserve the number of excitations, oscillate rapidly and can hence be discarded. The interaction Hamiltonian in the interaction picture then reads:
\begin{equation}
\label{eq::Atom:HwwI}
H_\text{int}^I(t) = -i\omega_A\sum_{ij}(\mathbf{A}^{(+)}\cdot\mathbf{D}_{ij}^{(-)}e^{-i\Delta t} - \mathbf{A}^{(-)}\cdot\mathbf{D}_{ij}^{(+)}e^{i\Delta t}), 
\end{equation}
with $\Delta = \omega_L - \omega_A$.
In literature, the expression $\mathbf{E}\cdot \mathbf{D}$ analogous to the potential energy of an electric dipole in an external electric field is usually used as the light-matter interaction, which appears to contradict Eq.~\refer{eq::Atom:HwwI}. At least in the case $\omega_L = \omega_A$ of resonant interaction this does not constitute a problem, however, as it holds that $\mathbf{E}^{(\pm)} = \pm i\omega_L \mathbf{A}^{(\pm)}$ and thus the two expressions coincide. 
The influence of a non-vanishing detuning can be made apparent by rewriting $\mathbf{E}^{(\pm)} = \pm i\omega_A(1 + \frac{\Delta}{\omega_A})\mathbf{A}^{(\pm)}$; and one observes that the difference is of order $\frac{\Delta}{\omega_A}$ and thus does not play a significant role, as long as the detuning is smaller than the energy difference of the atomic levels. 
The question of which approach is in principle the correct one shall not be pursued further in this work, instead we refer the reader to the discussion in Ref.~\cite{oregonscript}. In the following we will keep using the formulation involving the vector potential $\mathbf{A}$. \\

It is known from the investigation of Rabi oscillations in two-level systems that the detuning $\Delta$ has a wide influence on the maximum achievable occupation of the excited level when starting from exclusive occupation of the ground state. The occupation number of the excited state is negligible in particular if $\Delta$ is larger than the resonant Rabi frequency, which we will assume for the remainder of this section. 
Hence it suggests itself to exclude the excited states from consideration and develop an effective model for the dynamics of the ground states only. To this end we examine scattering amplitudes of the form 
$\braket{a(0)|b(t)} = \Braket{a(0)|\mathcal{T}\exp\left[-i\int_0^t\, H_\text{int}^I(\tau)d\tau\right]|b(0)}$ between ground states $\ket{a}$ and $\ket{b}$ in the interaction picture.

\begin{widetext}
The time evolution operator can be expanded into a Dyson series: 
\begin{equation}
\braket{a(0)|b(t)} = \Braket{a(0)|\mathds{1} - i\int_0^t\,  H_\text{int}^I(\tau) - \int_0^t\int_0^\tau\,  H_\text{int}^I(\tau) H_\text{int}^I(\tau^\prime)d\tau^\prime d\tau + ...|b(0)},
\end{equation}
whose zeroth and first order contribute only  $\delta_{ab}$, as the matrix elements of the interaction Hamiltonian vanish between two ground states. 

For the second-order term one finds
\begin{eqnarray*}
 H_\text{int}^I(\tau) H_\text{int}^I(\tau^\prime) &= & (-i\omega_A)^2\sum_{ijkl}\left(\mathbf{A}^{(+)}\cdot\mathbf{D}_{ij}^{(-)}\mathbf{A}^{(+)}\cdot\mathbf{D}_{kl}^{(-)}e^{-i\Delta(\tau + \tau^\prime)} - \mathbf{A}^{(+)}\cdot\mathbf{D}_{ij}^{(-)}\mathbf{A}^{(-)}\cdot\mathbf{D}_{kl}^{(+)}e^{-i\Delta(\tau - \tau^\prime)}\right.\\
& &\left. - \mathbf{A}^{(-)}\cdot\mathbf{D}_{ij}^{(+)}\mathbf{A}^{(+)}\cdot\mathbf{D}_{kl}^{(-)}e^{i\Delta(\tau - \tau^\prime)} + \mathbf{A}^{(-)}\cdot\mathbf{D}_{ij}^{(+)}\mathbf{A}^{(-)}\cdot\mathbf{D}_{kl}^{(+)}e^{i\Delta(\tau + \tau^\prime)}\right),
\end{eqnarray*}
of which when taking the matrix element between ground states only the third summand remains:
\begin{equation*}
\Braket{a(0)| - \mathbf{A}^{(-)}\cdot\mathbf{D}_{ij}^{(+)}\mathbf{A}^{(+)}\cdot\mathbf{D}_{kl}^{(-)}e^{i\Delta(\tau - \tau^\prime)}|b(0)} = \mathbf{A}^{(-)}\cdot\braket{i|\mathbf{D}|e_j}\mathbf{A}^{(+)}\cdot\braket{e_l|\mathbf{D}|k}\delta_{ai}\delta_{bk}\delta_{jl}e^{i\Delta (\tau-\tau^\prime)}.
\end{equation*}
The three Kronecker symbols collapse the four summations into one, resulting in the expression
\begin{equation}
\label{eq::Atom:streuamp}
\braket{a(0)|b(t)} = \delta_{ab} - \int_0^t\int_0^\tau\, \omega_A^2\sum_j\mathbf{A}^{(-)}\cdot\braket{a|\mathbf{D}|e_j}\mathbf{A}^{(+)}\cdot\braket{e_j|\mathbf{D}|b}e^{i\Delta(\tau-\tau^\prime)}d\tau^\prime d\tau
\end{equation}
for the scattering amplitude. 
\end{widetext}

Next we wish to interpret this amplitude as the result of an effective interaction Hamiltonian and hence make the ansatz $\braket{a(0)|b(t)} = \Braket{a(0)|\mathcal{T}\exp\left[-i\int_0^t \, H_\text{eff int}^I(\tau)d\tau\right]|b(0)}$. Expanding this in turn into a Dyson series, the $\delta_{ab}$ is reproduced by the zeroth order term while the non-trivial part of Eq.~\refer{eq::Atom:streuamp} should arise from the first-order term. Since the first-order term contains only one integral, however, the $\tau^\prime$-integral in Eq.~\refer{eq::Atom:streuamp} must be solved before a comparison is possible. Eventually one finds
\begin{eqnarray}
\label{eq::Atom:Heff}
H_\text{eff int}^I(t) &= &\sum_{abj}\omega_A^2\frac{\mathbf{A}^{(-)}\cdot\braket{a|\mathbf{D}|e_j}\mathbf{A}^{(+)}\cdot\braket{e_j|\mathbf{D}|b}}{\Delta}\nonumber\\
&&\times(1-e^{i\Delta t})\ket{a}\bra{b}
\end{eqnarray}
for the effective interaction. 
This Hamiltonian being time-dependent, a time-integration is required to obtain the corresponding time evolution operator, transforming the factor $(1 - e^{i\Delta t})$ into $(t - \frac{e^{i\Delta t} - 1}{i\Delta})$. Aside from the linear increase in $t$ a correction oscillating at frequency $\Delta$ becomes manifest, whose amplitude is suppressed by $1/\Delta$ though. If one disregards this correction, Eq.~\refer{eq::Atom:Heff} coincides with the result presented in Ref.~\cite{hammerer}, aside from the different formulations based on $\mathbf{E}$ and $\mathbf{A}$, respectively, and the conflicting definitions of $\Delta$ with opposite sign.

\subsection{Faraday interaction}
\begin{figure}
\includegraphics[width=0.4\textwidth]{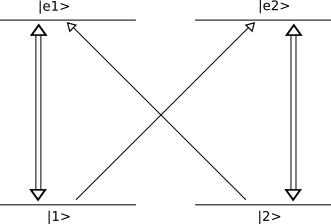}
\caption{Level scheme of Faraday interaction \cite{hammerer}. Two ground states $\ket{1}$ and $\ket{2}$ and two excited states $\ket{e_1}$ and $\ket{e_2}$ are coupled via two light modes. The first mode (double arrows) will be treated classically due to high intensity, coupling states with equal magnetic quantum number. The second mode (single arrows) is treated quantum mechanically and couples states with $\Delta m = \pm 1$.}
\label{fig::Far:levelscheme}
\end{figure}

After having derived an effective interaction between the ground states in the previous section, we now want to apply this to concrete level schemes. In particular we will consider the three cases presented in Ref.~\cite{hammerer}, starting with the Faraday interaction shown in Fig.~\ref{fig::Far:levelscheme}. For reasons of compatibility with Ref.~\cite{hammerer} we assume the quantization axis to be aligned with the $x$-direction. \\

As the structure of the atomic energy levels we assume the comparatively simple case of two ground states and two excited states, that can accordingly be understood as magnetic sublevels with the quantum numbers $m = - \frac{1}{2}$ for $\ket{1}$ and $m = \frac{1}{2}$ for $\ket{2}$ to the angular momentum $j = \frac12$. As apparent from Fig.~\ref{fig::Far:levelscheme}, all possible transitions between ground- and excited states are supposed to be coupled by two electromagnetic modes, one of which stimulates vertical transition with $\Delta m = 0$ and is treated classically due to high intensity, while the other stimulates cross transitions with $\Delta m = \pm 1$ and is treated as a quantum mode. 

In order to be able to stimulate $\Delta m = 0$-transitions the classical mode has to be linearly polarized in $x$-direction, and thus propagate in the $y$-$z$-plane. Cross transitions require circular polarization in the $y$-$z$-plane, albeit with opposite orientation for  $\Delta m = +1$ and $\Delta m = -1$, thus a mode meant to couple to both transitions must be linearly polarized in the $y$-$z$-plane, allowing the polarization to be considered the sum of both circular components. The Faraday interaction as described in Refs.~\cite{hammerer} and \cite{lightgkp} is generated using $y$-polarization, hence we examine this case first. \\

Calculating the interaction Hamiltonian for a concrete level scheme requires evaluating the dipole operator's matrix elements. This is facilitated by the Wigner-Eckart theorem 
\begin{equation}
\braket{\alpha^\prime, j^\prime m^\prime|T^{(k)}_q|\alpha, jm} = \braket{jm;kq|j^\prime m^\prime}\braket{\alpha^\prime, j^\prime||T^{(k)}||\alpha, j},
\end{equation}
with  $ -k \leq q \leq k$. It ascribes the entire $m$- and $M^\prime$-dependency to a Clebsch-Gordan coefficient, while the reduced matrix element $\braket{\alpha^\prime, j^\prime||T^{(k)}||\alpha, j}$ only depends on $j$ and $j^\prime$, as well as possibly other not angular momentum related quantum numbers. The Clebsch-Gordan coefficient also gives rise to the selection rules regarding polarization mentioned earlier, as it is only non.zero for  $m + q = m^\prime$.
The dipole operator as a vector operator is a spherical tensor of rank $k=1$, and its spherical components are given by
\begin{equation}
D_{\pm1} = \mp\frac{D_y \pm i D_z}{\sqrt{2}} ,\qquad D_0 = D_x
\end{equation}
if the quantization axis is along $x$.
 The non-vanishing matrix elements of the dipole operator thus read
\begin{eqnarray}
\braket{1|D_0|e_1} &\propto &\braket{\frac12, -\frac12; 1, 0|\frac12, -\frac12} = -\sqrt{\frac13} \nonumber\\
\braket{1|D_{-1}|e_2} &\propto &\braket{\frac12, \frac12; 1, -1|\frac12, -\frac12} = \sqrt{\frac23} \nonumber\\
\braket{2|D_{+1}|e_1} &\propto &\braket{\frac12, -\frac12; 1, 1|\frac12, \frac12} = -\sqrt{\frac23} \nonumber\\
\braket{2|D_0|e_2} &\propto &\braket{\frac12, \frac12; 1, 0|\frac12, \frac12} = \sqrt{\frac13}
\label{eq::Far:dipolelements}
\end{eqnarray}
and 
\begin{eqnarray}
\braket{e_1|D_0|1} &\propto &\braket{\frac12, -\frac12; 1, 0|\frac12, -\frac12} = -\sqrt{\frac13} \nonumber\\
\braket{e_2|D_{+1}|1} &\propto &\braket{\frac12, -\frac12; 1, -1|\frac12, \frac12} = -\sqrt{\frac23} \nonumber\\
\braket{e_1|D_{-1}|2} &\propto &\braket{\frac12, \frac12; 1, 1|\frac12, -\frac12} = \sqrt{\frac23} \nonumber\\
\braket{e_2|D_0|2} &\propto &\braket{\frac12, \frac12; 1, 0|\frac12, \frac12} = \sqrt{\frac13},
\label{eq::Far:dipolelements2}
\end{eqnarray} 
with the reduced matrix elements $\braket{g, \frac12||D||e, \frac12}$ in Eq.~\refer{eq::Far:dipolelements} and its complex conjugate $\braket{e, \frac12||D||g, \frac12}$ in Eq.~\refer{eq::Far:dipolelements2}, abbreviated in the following as $\overline{D}$ and $\overline{D}^\ast$  respectively, as proportionality constants. \\

The scalar product of the polarization vectors with the dipole operator takes the form $\bm{\epsilon} \cdot \mathbf{D} = \sum_q (-1)^q \epsilon^\ast_qD_{-q}$ in spherical coordinates. The relevant polarization components are $\epsilon^\text{classical}_0 = \epsilon^\text{classical}_x = 1$ for the classical field and $\epsilon^\text{quantum}_{\pm 1} = \mp(\epsilon^\text{quantum}_y \pm i\epsilon^\text{quantum}_z)/\sqrt{2} = \mp\frac{1}{\sqrt{2}}$ for the quantum field. Additionally, both polarization vectors are real, so that no distinction between $\bm{\epsilon}$ and $\bm{\epsilon}^\ast$ is necessary.  Using this information, the matrix elements can be evaluated by inserting Eqs.~\refer{eq::Far:dipolelements} and \refer{eq::Far:dipolelements2} into 
\begin{equation}
\braket{a|H_{\text{int}}|b} = \frac{\omega_A^2}{\Delta}\sum_j \mathbf{A}^{(-)}\cdot\braket{a|\mathbf{D}|e_j}\mathbf{A}^{(+)}\cdot\braket{e_j|\mathbf{D}|b},
\label{eq::Far:Vab}
\end{equation}
where the sum runs over excited states. $\mathbf{A}$ theoretically consists of a sum of the classical- and the quantum field in this equation, but the scalar product filters out unwanted contributions, such that the classical field only plays a role for vertical transitions and the quantum field only for cross transitions. The Hamiltonian's matrix elements now read
\begin{eqnarray}
\braket{1|H_{\text{int}}|1} &= &\frac{\omega_A^2|\overline{D}|^2|C|^2}{\Delta}\left(\frac13|\alpha|^2 + \frac13a^\dagger a\right) \nonumber\\
\braket{1|H_{\text{int}}|2} &= &\frac{\omega_A^2|\overline{D}|^2|C|^2}{\Delta}\left(- \frac13\alpha^\ast a + \frac13 \alpha a^\dagger\right) \nonumber\\
\braket{2|H_{\text{int}}|1} &= &\frac{\omega_A^2|\overline{D}|^2|C|^2}{\Delta}\left(- \frac13 \alpha a^\dagger + \frac13 \alpha^\ast a\right) \nonumber\\
\braket{2|H_{\text{int}}|2} &= &\frac{\omega_A^2|\overline{D}|^2|C|^2}{\Delta}\left(\frac13 a^\dagger a + \frac13|\alpha|^2\right),
\end{eqnarray}
where $\alpha$ is the amplitude of the classical field, which is accordingly written as $\mathbf{A} = C(\bm{\epsilon}^\text{classical}\alpha + \bm{\epsilon}^{\ast\text{classical}}\alpha^\ast)$.\\

In order to extend to an entire ensemble of atoms characterized by the level scheme in Fig.~\ref{fig::Far:levelscheme}, the Hamiltonian merely needs to be summed over all $N$ atoms:
\begin{eqnarray}
H_\text{int Ens} &= &\sum_{i=1}^N H_\text{int} = \sum_{i=1}^N \sum_{ab} \braket{a|H_\text{int}|b}\ket{a}\bra{b} \nonumber\\
&= &N\frac{\omega_A^2|\overline{D}|^2|C|^2}{\Delta}\left(\frac13|\alpha|^2 + \frac13a^\dagger a\right) \\
&&- \frac{\omega_A^2|\overline{D}|^2|C|^2}{\Delta}\left(\frac13 \alpha a^\dagger - \frac13 \alpha^\ast a\right)(J_+ - J_-).\nonumber
\end{eqnarray}
The first summand does not contain operators of the ensemble and can thus be transferred from the interaction to the free Hamiltonian. Physically, it can be interpreted as a refractive index of the ensemble experienced by the light \cite{hammerer}.  
In the second summand $a^\dagger - a$ can be rewritten as $-\sqrt{2}ip$ and $J_+ - J_-$ as $2iJ_z$, and one obtains, if $\alpha$ is assumed real,  $H_\text{WW} \propto -pJ_z$ with proportionality constant $\frac{2\sqrt{2}\omega_A^2|\overline{D}|^2|C|^2\alpha}{3\Delta}$ for the interaction Hamiltonian. This interaction is the basis for the creation of GKP states in an electromagnetic mode as described in Ref.~\cite{lightgkp}. 

In Holstein-Primakoff approximation the Hamiltonian is proportional to $-pP$, where lower case letters still denote quadratures of the field and upper case letters those of the ensemble, and the sign can be chosen as desired by adjusting $\Delta$ to be either red- or blue-detuned. 
In principle, it is also possible to generate an interaction proportional to $pJ_y$ or $pX$ by choosing the polarization of the quantum mode along $z$ instead of along $y$. 

\begin{figure*}
\begin{minipage}{0.45\textwidth}
\includegraphics[width=0.9\textwidth]{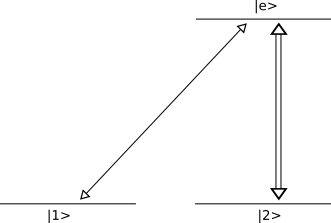}
\end{minipage}
\vrule
\begin{minipage}{0.45\textwidth}
\includegraphics[width=0.9\textwidth]{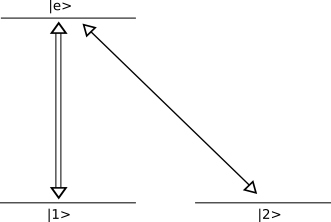}
\end{minipage}
\caption{Level scheme of beam-splitter (left) and squeezing interaction (right) \cite{hammerer}. Two ground states $\ket{1}$ and $\ket{2}$, one excited state $\ket{e}$ having $m = +1/2$ (left) bzw. $m = -1/2$ (right) are coupled via two light modes. The classical mode couples $\Delta m = 0$ - transitions from $\ket{2}$ to $\ket{e}$ or $\ket{1}$ to $\ket{e}$, respectively; the quantum mode couples either to $\Delta m = 1$ or $\Delta m = -1$ - transitions, depending on polarization.}
\label{fig::ST:levelscheme}
\end{figure*}

\begin{figure}
\includegraphics[width=0.45\textwidth]{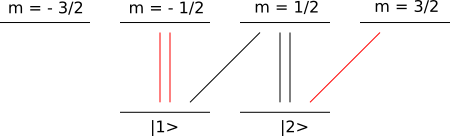}
\caption{Extended level scheme for the beam-splitter interaction. Four excited states enable new transitions (red) for unchanged field configuration, affecting the diagonal elements of the Hamiltonian. The non-diagonal elements remain unchanged. }
\label{fig::ST:levelextended}
\end{figure}

\subsection{Beam-splitter and Squeezing interaction}
 Next we turn to beam arrangements that enable simulating beam splitters and two-mode squeezing gates between a light mode and an atomic ensemble in Holstein-Primakoff approximation. The corresponding level schemes are depicted in Fig.~\ref{fig::ST:levelscheme}. As was the case for the Faraday interaction, two light modes are used, one of which is treated classically and the other quantum mechanically. The classical mode is still linearly polarized along the quantization axis $x$ and enables transitions of equal magnetic quantum number, whereas the orientation of the quantum mode's circular polarization decides whether a beam splitter or squeezing interaction is obtained: in $\sigma_+$ configuration it couples the ground state $\ket{1}$ with $m = -\frac12$ with the excited state having $m = \frac12$ and thus creates an atomic spin excitation while absorbing a photon and vice versa, leading to the expectation of $aJ_+ + a^\dagger J_-$ as the Hamiltonian, which would correspond to a beam splitter interaction. In contrast, $\sigma_-$-polarized light either creates or destroys both a photon and a spin excitation, corresponding to a two-mode squeezing. \\

To calculate the interactions, we start by again considering the spherical components of the polarization vectors, starting with $\bm{\epsilon} = \begin{pmatrix}0 & 1 & -i\end{pmatrix}^T/\sqrt{2}$ for the beam splitter interaction. One finds: 
\begin{eqnarray}
\epsilon^\text{quantum}_{+1} &= &-\frac{1}{2}(1 + i(-i)) = -1\nonumber\\
\epsilon^\text{quantum}_{-1} &= &\frac{1}{2}(1 - i(-i)) = 0\nonumber\\
\epsilon^\text{quantum}_0 &= &0,
\end{eqnarray}
and for the complex conjugated vector:
\begin{eqnarray}
(\epsilon^\ast)^\text{quantum}_{+1} &= &-\frac{1}{2}(1 + i^2) = 0\nonumber\\
 (\epsilon^\ast)^\text{quantum}_{-1} &= &\frac{1}{2}(1 - i^2) = 1\nonumber\\
(\epsilon^\ast)^\text{quantum}_0 &= &0.
\end{eqnarray}

Together with Eqs.~\refer{eq::Far:dipolelements} and \refer{eq::Far:dipolelements2} follows
\begin{eqnarray}
\braket{1|H_{\text{int}}|1} &= \frac{\omega_A^2|\overline{D}|^2|C|^2}{\Delta}\frac23a^\dagger a \nonumber\\
\braket{1|H_{\text{int}}|2} &= \frac{\omega_A^2|\overline{D}|^2|C|^2}{\Delta}\frac{\sqrt{2}}{3} \alpha a^\dagger \nonumber\\
\braket{2|H_{\text{int}}|1} &= \frac{\omega_A^2|\overline{D}|^2|C|^2}{\Delta} \frac{\sqrt{2}}{3} \alpha^\ast a \nonumber\\
\braket{2|H_{\text{int}}|2} &= \frac{\omega_A^2|\overline{D}|^2|C|^2}{\Delta}\frac13|\alpha|^2
\end{eqnarray}
for the effective interaction. 
In contrast to the Faraday interaction, there is only one path for each transition between ground states, resulting in the matrix elements no longer consisting of two summands. 

As a side effect of this, the diagonal elements  $\braket{1|H_{\text{int}}|1}$ and $\braket{2|H_{\text{int}}|2}$  are no longer identical and one does not obtain a term of the form $\braket{1|H_{\text{int}}|1}(\ket{1}\bra{1} + \ket{2}\bra{2}) = \braket{1|H_{\text{int}}|1}\mathds{1}$ in the final Hamiltonian, as was the case before, and the diagonal contributions cannot be simply transferred to the free part. Thus, it appears as if the resulting interaction did not only contain the expected beam splitter structure, but additional terms arising from the diagonal. 

This problem can be solved by admitting more than one excited state, as illustrated in Fig.~\ref{fig::ST:levelextended}. The set of excited states can e.g. be considered as a spin $\frac34$ - system with 4 magnetic sublevels, which is a more realistic scenario anyway due to the selection rule $\Delta j = \pm 1$ for optical dipole transitions. 
The classical field then additionally couples the $\Delta m = 0$ transition between $\ket{1}$ and the new $m = -\frac12$ state, leading to a term proportional to $|\alpha|^2$ in $\braket{1|H_{\text{int}}|1}$.
The quantum field enables a new transition from $\ket{2}$ to the excited state with $m = \frac32$, producing an $a^\dagger a $ term in $\braket{2|H_{\text{int}}|2}$. Even though the Clebsch-Gordan coefficients cause a slight deviation between the two new diagonal elements, they can now be approximated as being equal, since the $a^\dagger a $ contributions are negligible compared to $|\alpha|^2$. 
 After this modification we again sum over all atoms in the ensemble and obtain $H_\text{int} \propto aJ_+ + a^\dagger J_-$ or, in Holstein-Primakoff approximation, $H_\text{int} \propto a_La_A^\dagger + a_L^\dagger a_A$.\\

In the case of squeezing interaction, the polarization vector reads $\bm{\epsilon} = \begin{pmatrix}0 & 1 & i \end{pmatrix}^T /\sqrt2$, i.e. the roles of $\bm{\epsilon}$ and $\bm{\epsilon}^\ast$ are switched compared to the previous case. When evaluating the matrix elements, one runs into a similar problem as before, which is in turn solved by extending the level scheme. As expected, one finally obtains the Hamiltonian $H_\text{int} \propto aJ_- + a^\dagger J_+$ respectively $H_\text{int} \propto a_La_A + a_L^\dagger a_A^\dagger$.
In the framework of a second-generation repeater, the two-mode squeezing interaction is essential to perform amplification in order to transform losses from memory storage into Gaussian shifts.

\section{Statistics of relevant random variables}
\subsection{Difference of two geometric random variables}
\label{subsec::appGeo}
Let $\mathcal{N}_1$ and $\mathcal{N}_2$ be two geometrically distributed random variables with success probability $p$. In the following we will derive the probability distribution and the expectation value of $|\mathcal{N}_1 - \mathcal{N}_2|$ (see also Refs.~\cite{frank_matfqkd, pvl_extendingqlinks}). 

At first we assume $\mathcal{N}_1 \geq \mathcal{N}_2$ and calculate the probability of the difference taking the value $k \geq 0$. This is the case if and only if for $\mathcal{N}_1 = n$, $\mathcal{N}_2$ equals $n-k$, thus
\begin{eqnarray}
P(\mathcal{N}_1 - \mathcal{N}_2 = k) &=  &\sum_{n = k+1}^\infty P(\mathcal{N}_1 = n)P(\mathcal{N}_2 = n-k) \nonumber\\
&= &\sum_{n = k+1}^\infty pq^{n-1}pq^{n-k-1}\nonumber\\
& = &\frac{p^2}{q^{2+k}}\sum_{n = k+1}^\infty (q^2)^n,
\label{eq::appGeo:kpos}
\end{eqnarray}
where $q = 1-p$ and the lower summation bound is chosen such that both $\mathcal{N}_1$ and $\mathcal{N}_2$ are valid realizations of a geometric random variable, i.e. $\mathcal{N}_{1, 2} \geq 1$. 
The sum in Eq.~\refer{eq::appGeo:kpos} is reminiscent of a geometric series; however, the lower bound is not 0 and the sum needs to be rewritten using $\sum_{n=k+1}^\infty = \sum_{n=0}^\infty - \sum_{n=0}^k$ before the well-known results about the limit and the finite partial sums of the geometric series can be applied. One then finds
\begin{equation}
P(\mathcal{N}_1 - \mathcal{N}_2 = k) =  \frac{p^2}{q^{2+k}} \left[\frac{1}{1-q^2} - \frac{1-q^{2k+2}}{1-q^2}\right] = \frac{p^2q^k}{1-q^2}.
\end{equation}
The bounds in Eq.~\refer{eq::appGeo:kpos} are only correct for non-negative $k$. The appropriate expression for $k < 0$ is
\begin{equation}
P(\mathcal{N}_1 - \mathcal{N}_2 = k) = \sum_{n = 1}^\infty pq^{n-1}pq^{n-k-1} ,
\end{equation}
since the condition $n-k \geq 1$ is always fulfilled anyway. The other steps are carried out as before and one obtains $P(\mathcal{N}_1 - \mathcal{N}_2 = k) = \frac{p^2q^{-k}}{1-q^2} $ for $k < 0$. \\

The distribution of the difference's absolute value can now be read off immediately:
\begin{equation}
\label{eq::appGeo:verteilung}
P(|\mathcal{N}_1 - \mathcal{N}_2| = k) = \begin{cases}\frac{p^2}{1-q^2} & k = 0\\
\frac{2p^2q^k}{1-q^2} & k \neq 0.\end{cases}
\end{equation}
 The expectation value $\mathds{E}(|\mathcal{N}_1 - \mathcal{N}_2|)$ can be calculated as well:
\begin{eqnarray}
\mathds{E}(|\mathcal{N}_1 - \mathcal{N}_2|) &= &\sum_{k = 0}^\infty k P(|\mathcal{N}_1 - \mathcal{N}_2| = k) = \frac{2p^2}{1-q^2}\sum_{k=1}^\infty kq^k \nonumber\\
 &= &\frac{2p^2}{1-q^2}\frac{q}{(q-1)^2} = \frac{2q}{1-q^2},
\end{eqnarray}
where use was made of the identity $\sum_{k = 0}^\infty kq^k = \frac{q}{(q-1)^2}$.

\subsection{Distribution of total waiting time}\label{waitingstatistics}

\begin{figure*}
\begin{minipage}{0.50\textwidth}
\includegraphics[width=\textwidth]{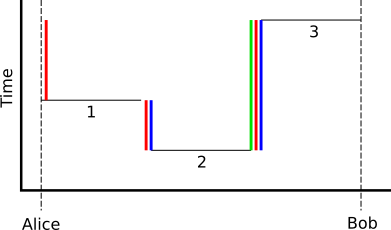}
\end{minipage}
\begin{minipage}{0.48\textwidth}
\includegraphics[width=\textwidth]{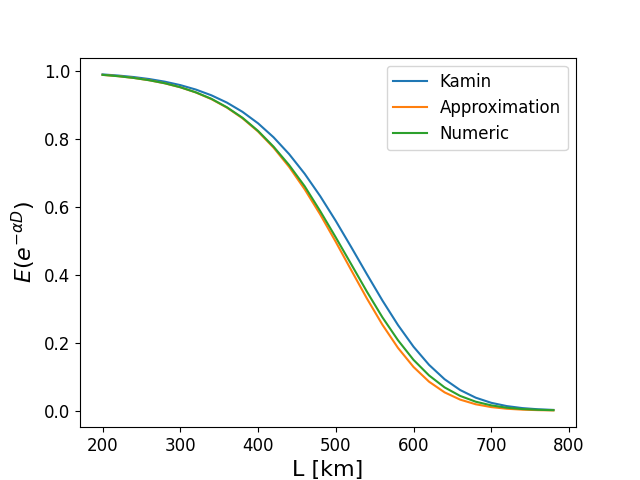}
\end{minipage}
\caption{Left: Illustration of total dephasing time in a three-segment repeater with temporal distribution order 2, 1, 3. The meaning of the lines: Red corresponds to the sum of all memory waiting times, including Alice's (and generally, for other temporal orders, Bob's) memories, Green is one half of Red and corresponds to the general definition used in Ref.~\cite{rateanalysis}, and Blue represents the definition used in our derivation for Eq.~(\ref{eq::Rate:kaminana}). Right: Average dephasing for four segments determined by the exact formula from Ref.~\cite{rateanalysis}, our approximation (Eq.~(\ref{eq::Rate:kaminana})), and a numerical simulation based on our definition of $D_n$. Due to the differing definitions, the blue and orange curves do not overlap, whereas the green and orange curves barely differ.}
\label{fig::dephasing_approx}
\end{figure*}
In this section, we will derive an approximate probability distribution for the total waiting time $D_n$ and the related exponential average $\mathds{E}(e^{-\alpha D_n})$, which is needed to calculate secret key rates for the correctionless scheme from Ref.~\cite{rateanalysis}
(in Ref.~\cite{rateanalysis}, $D_n$ is referred to as the total dephasing time). 
Approximate here refers to the fact that we will make the assumption that
the total waiting time can be understood as the sum of $n-1$ independent random variables $X_i$ distributed according to Eq.~\refer{eq::appGeo:verteilung}. 
However, more precisely, the $n-1$ absolute values of the differences of two geometrically distributed random variables are actually not statistically independent. Nonetheless, for the parameter values that are of interest to us, this plays a negligible role for the average dephasing, as can be seen in Fig.~\ref{fig::dephasing_approx}, where the difference between the green and orange curves is barely visible. 

Aside from the assumption of independent random variables, calculating the effect of the total memory loss based on the total average memory waiting times accumulated at every one of the $n-1$ intermediate repeater stations, excluding the sender (Alice) and the receiver (Bob) stations, relies on the following basic assumptions \cite{rateanalysis}: 
(i) deterministic entanglement swapping, (ii) swapping as soon as possible, and (iii) Alice and Bob measuring their atoms immediately in a QKD application, thus freeing them of any storage-induced loss errors occurring at the repeater's sender and receiver stations. 
Note that assumption (iii) means that our definition of $D_n$ differs from the general definition as used in Ref.~\cite{rateanalysis}, where $D_n$ is one half the time spent waiting by all the memories in the chain and where Alice and Bob's memories generally store their qubits until the time of the final swapping
(though Ref.~\cite{rateanalysis} does make the distinction between the cases where Alice and Bob measure immediately and where they do not, and some examples are also presented for the former case). 

The differences between these definitions can be best understood by considering the simple case of 3 segments, illustrated in Fig.~\ref{fig::dephasing_approx}.
If the segments finish distribution in the order $2, 1, 3$, our definition will include the edge between segments 1 and 2 as well as between 2 and 3 (blue lines). If one does not consider immediate measurements by Alice and Bob, the edge from segment 1 up to the finishing time of segment 3 must be included (red lines). In these cases, at any point in time there are two memories that are waiting. Considering always only one memory waiting (green line) corresponds to a coherence time twice as large as for the more general case of two memories subject to decoherence, which then could be easily adapted by redefining the coherence time \cite{rateanalysis}.
Unlike the scenario where only one memory experiences dephasing at any time, in our convention, two memories are  sometimes subject to loss (and Pauli noise on the logical level) simultaneously, such that at equal coherence times, our expectation value $\mathds{E}(e^{-\alpha D_n})$ will be lower than that of Ref.~\cite{rateanalysis} (see Fig.~\ref{fig::dephasing_approx} right). Note that even when the first distribution attempt in a repeater segment is successful, typically the corresponding memories must wait until the photons have been sent to the neighboring detectors and until the classical signals come back to them. In Ref.~\cite{rateanalysis}, this constant minimal, initial dephasing has been neglected in several cases, whereas in our calculations and plots, including the comparisons with Ref.~\cite{rateanalysis}, it has always been included for both types of schemes -- with corrections and correctionless.\\

Now we derive the distribution of a sum $S$ having $m$ summands that are independently distributed according to Eq.~\refer{eq::appGeo:verteilung} (the letter $S$ for this sum is only used in this subsection, while in the rest of the paper  $S$ stands for the secret key rate). \\

Obviously it holds that
\begin{equation}
P(S = j) = \sum_{k_1+...+k_m = j} P(X_1 = k_1)\cdots P(X_m = k_m).
\end{equation}
For $j = 0$ the sum contains only one term, as all $m$ random variables have to take the value 0, so that one can immediately state $P(S = 0) = \left(\frac{p^2}{1-q^2}\right)^m$. For $j \neq 0$ the situation is more complex, particularly since due to the form of Eq.~\refer{eq::appGeo:verteilung} one needs to distinguish between variables taking the value 0 vs. non-zero values. However, every summand contains the expression $\left(\frac{p^2}{1-q^2}\right)^mq^j$ that can therefore be factored out. For every variable taking a non-zero value one additionally obtains a factor 2. 

Assume we have a term where $i$ random variables are non-zero. Then there are $\begin{pmatrix}m\\i\end{pmatrix}$ possibilities to distribute them among the $m$ variables. We now have to find the number of combinations of $i$ strictly positive natural numbers with sum $j$, which is facilitated by the following geometrical consideration:
Imagine every possible combination as a point in an $i$-dimensional space. Points with a certain 1-norm lie in a hyperplane orthogonal to the vector (1, 1, ..., 1), which means in particular that any two points differ in more than one component. Therefore one can choose a coordinate and project into the hyperplane orthogonal to it, without changing the number of points, as this projection will be injective. The points now form a $i-1$-dimensional simplex with $j - i + 1$ points along each edge. 
The number of points can thus be extracted using the formula for figurate numbers and it reads
\begin{equation}
\begin{pmatrix}(j-i+1) + (i-1) - 1\\i-1\end{pmatrix} = \begin{pmatrix}j-1\\i-1\end{pmatrix}.
\end{equation}

Overall the probability for $j \neq 0$ is therefore given by
\begin{equation}
P(S = j) = \left(\frac{p^2}{1-q^2}\right)^mq^j\sum_{i=1}^m 2^i   \begin{pmatrix}m\\i\end{pmatrix}\begin{pmatrix}j-1\\i-1\end{pmatrix}.
\end{equation}\\

Next we turn to the expectation value $\mathds{E}(e^{-\alpha S})$. Using the distribution of $S$ we can write
\begin{widetext}
\begin{equation}
\mathds{E}(e^{-\alpha S}) = \left(\frac{p^2}{1-q^2}\right)^m +  \sum_{j=1}^\infty e^{-\alpha j}\left(\frac{p^2}{1-q^2}\right)^mq^j\sum_{i=1}^m 2^i   \begin{pmatrix}m\\i\end{pmatrix}\begin{pmatrix}j-1\\i-1\end{pmatrix}
\end{equation}
or, after rewriting and performing an index shift
\begin{equation}
\mathds{E}(e^{-\alpha S}) =  \left(\frac{p^2}{1-q^2}\right)^m + \left(\frac{p^2}{1-q^2}\right)^m\sum_{i=1}^m \begin{pmatrix}m\\i\end{pmatrix}2^i\sum_{j=0}^\infty (qe^{-\alpha})^{j+1} \begin{pmatrix}j\\i-1\end{pmatrix}.
\end{equation}
The $j$-summation can be performed using the identity $\sum_{j=0}^\infty \begin{pmatrix}j\\k\end{pmatrix} \frac{1}{z^{j+1}} = \frac{1}{(z-1)^{k+1}}$ with $z = \frac{1}{qe^{-\alpha}}$ and one obtains
\begin{equation}
\mathds{E}(e^{-\alpha S}) =  \left(\frac{p^2}{1-q^2}\right)^m + \left(\frac{p^2}{1-q^2}\right)^m\sum_{i=1}^m \begin{pmatrix}m\\i\end{pmatrix}2^i\left(\frac{qe^{-\alpha}}{1 - qe^{-\alpha}}\right)^i.
\end{equation}

The $i$-sum bears similarity to the right-hand side of the binomial theorem, so that after substituting the left-hand side we find
\begin{equation}
\mathds{E}(e^{-\alpha S}) =  \left(\frac{p^2}{1-q^2}\right)^m +  \left(\frac{p^2}{1-q^2}\right)^m \left[\left(1 + \frac{2qe^{-\alpha}}{1 - qe^{-\alpha}}\right)^m - 1\right] 
 =  \left(\frac{p^2}{1-q^2}\right)^m \left(\frac{1 + qe^{-\alpha}}{1 - qe^{-\alpha}}\right)^m.
\end{equation} 
\end{widetext}

This corresponds to Eq.~(\ref{eq::Rate:kaminana}) replacing $D_n$ for the general sum $S$ and $n-1$ stations for the general number of summands $m$.

\subsection{Average Pauli error probability}
For our rate analysis, we approximate the average Pauli error probability for each swapping as the probability corresponding to the average variance, i.e.
$\mathds{E}[p_\text{Pauli}(2\delta^2 + \sigma_\text{add}^2)] \approx p_\text{Pauli}(\mathds{E}[2\delta^2 + \sigma_\text{add}^2])$, with $p_\text{Pauli}$ defined in Eq.~\refer{eq::Rate:ppauli}. 
Here we want to estimate the error of this approximation. 

To do this, we expand $p_\text{Pauli}$ as a function of $\sigma_\text{add}^2$ around the natural variance $2\delta^2$. The true average is then given by
\begin{widetext}
\begin{eqnarray}
\mathds{E}[p_\text{Pauli}(2\delta^2 + \sigma_\text{add}^2)] &= &\sum_{k=0}^\infty P(k) \left[p_\text{Pauli}(2\delta^2) + p_\text{Pauli}^\prime(2\delta^2)\sigma_\text{add}^2(k) +\frac12 p_\text{Pauli}^{\prime\prime}(2\delta^2)(\sigma_\text{add}^2(k))^2 + ... \right]\\
& = & p_\text{Pauli}(2\delta^2) +  p_\text{Pauli}^\prime(2\delta^2)\mathds{E}[\sigma_\text{add}^2] + \frac12 p_\text{Pauli}^{\prime\prime}(2\delta^2)\mathds{E}[\sigma_\text{add}^4] + ... ,
\end{eqnarray}
where $P(k)$ is the probability distribution of the waiting time derived in Section~\ref{subsec::appGeo}. 
In our approximation, we replaced this with
\begin{equation}
 p_\text{Pauli}(\mathds{E}[2\delta^2 + \sigma_\text{add}^2]) = p_\text{Pauli}(2\delta^2) +  p_\text{Pauli}^\prime(2\delta^2)\mathds{E}[\sigma_\text{add}^2] + \frac12 p_\text{Pauli}^{\prime\prime}(2\delta^2)\mathds{E}[\sigma_\text{add}^2]^2 + ...
\end{equation}
\end{widetext}
The difference is thus of the order 
\begin{equation}
\frac12 p_\text{Pauli}^{\prime\prime}(2\delta^2)(\mathds{E}[\sigma_\text{add}^4]  - \mathds{E}[\sigma_\text{add}^2]^2) = \frac12 p_\text{Pauli}^{\prime\prime}(2\delta^2)\text{Var}[\sigma_\text{add}^2].
\end{equation}
While $p_\text{Pauli}$ has no closed form expression, its derivatives do, and the second derivative is given by
\begin{equation}
p_\text{Pauli}^{\prime\prime}(\sigma^2) = \frac{1}{\sqrt{8}} e^{-\frac{\pi}{8\sigma^2}}\left(\frac{\pi}{8}\frac{1}{\sigma^7} - \frac32\frac{1}{\sigma^5}\right),
\end{equation}
while the variance in the case of preamplification reads
\begin{equation}
\text{Var}[\sigma_\text{add}^2] = (1 - e^{-\alpha})^2\left(\frac{2q}{(1-q)^2} - \frac{4q^2}{(1-q^2)^2}\right).
\end{equation}
The error can be seen to be strongly depended on $\delta^2$ and $q$; however, at the parameter values considered in this work, the relative error is usually in the order of $10^{-2}$ or lower, whenever a non-zero rate is obtained.

\subsection{Expected variance of CC-amplification}\label{CCexpect}
The additional variance caused by CC-amplification is given in Eq.~\refer{eq::AS:CC1} as 
\begin{equation*}
\sigma_\text{add}^2 = \frac{1-e^{-(t_\text{wait}+1)\alpha}}{e^{-(t_\text{wait}+1)\alpha}}.
\end{equation*}
Here we derive the expectation value based on the distribution of the waiting time for a single segment as presented in Section~\ref{subsec::appGeo}. 
We have
\begin{eqnarray}
\mathds{E}(\sigma_\text{add}^2) &= &\frac{p^2}{1-q^2} \frac{1 - e^{-\alpha}}{e^{-\alpha}} + \sum_{N=1}^\infty \frac{2p^2}{1-q^2} q^N\frac{1-e^{-\alpha(N+1)}}{e^{-\alpha(N+1)}} \nonumber\\
&= &\frac{p^2}{1-q^2} \frac{1 - e^{-\alpha}}{e^{-\alpha}} + \sum_{N=1}^\infty \frac{2p^2}{1-q^2} q^N \left(e^{\alpha(N+1)} - 1\right) \nonumber\\
&= &\frac{p^2}{1-q^2} \left(\frac{1 - e^{-\alpha}}{e^{-\alpha}} \right.\nonumber\\
&&+  \left.2e^{\alpha}\sum_{N=1}^\infty (e^\alpha q)^N - 2\sum_{N=1}^\infty q^N\right).\quad
\end{eqnarray}
Assuming $q < e^{-\alpha}$ we can make use of the geometric series limit and write

\begin{equation}
\mathds{E}(\sigma_\text{add}^2) = \frac{p^2}{1-q^2} \left[\frac{1 - e^{-\alpha}}{e^{-\alpha}} + \frac{2qe^{2\alpha}}{1 - qe^\alpha} - \frac{2q}{1-q}\right],
\end{equation}
thus arriving at Eq.~\refer{eq::AS:CC}.

\section{List of symbols}
\label{listofsymbols}
In this section we provide a list of the variables used in the main text and the appendices of this paper.\\

\begin{table*}
\begin{ruledtabular}
\begin{tabular}{ll}
Symbol \hphantom{m}& Meaning\\
\hline
&\\
&\textbf{\textit{Constants}}\\
&\\
$c $ & Speed of light in optical fiber (2 $\times 10^{5}$km/s )\\
$L_\text{att} $ & Attenuation length of optical fiber at telecom wavelength (22km)\\
&\\

& \textbf{\textit{Variables in main text}}\\
&\\
$L$ & Total length of repeater chain\\
$L_0$ & Length of each segment \\
$n$ & Number of segments\\
$N$ & Number of atoms in ensemble\\
$p$ & Probability of entanglement distribution success in one segment\\
$q$ & $1 - p$\\
$p_\text{link}$ & Link efficiency in one segment\\
$p_\text{Pauli}$ & Probability of Pauli error occurring in one swapping event\\
$\eta$ & Transmissivity of memory loss channel\\
$\alpha^\prime$ & Inverse coherence time (dimension time$^{-1}$)\\
$\alpha$ & Effective inverse coherence time (dimensionless)\\
$\tau$ & Duration of one timestep due to classical communication\\
$t_\text{coh}$ & Memory coherence time\\
$t_\text{wait}$ & Number of timesteps a memory of a finished segment spends waiting for the neighbouring segment (Random variable) \\
$T$ & $\mathds{E}(t_\text{wait})$\\
$\delta^2$ & Finite GKP squeezing\\
$\Delta^2$ & Variance of envelope for realistic GKP states\\
$\gamma^2$ & Variance introduced by imperfect gates or measurements\\
$\sigma^2_\text{add}$ & Variance affecting Bell measurement when swapping that is not due to finite GKP squeezing (Random variable)\\
$\sigma^2_\text{Bell}$ & Variance of Bell pair in imperfect teleportation\\
$\sigma^2_\text{tot}$ & Total variance affecting Bell measurement\\
$D_n$ & Sum of all timesteps that memories spend waiting (Random variable)\\
$\overline{K_n}$ & Average number of timesteps until Alice and Bob share a Bell pair \\
$R$ & Raw rate\\
$r$ & Secret key fraction\\
$S$ & Secret key rate\\
&\\

& \textbf{\textit{Variables in the appendices}}\\
&\\
$N$ & Number of atoms in ensemble\\
$\mathcal{N}$ & Geometric random variable (including successful attempt)\\
$m_e$ & Electron mass\\
$\mathbf{A}$ & Electromagnetic vector potential (operator)\\
$\tilde{\alpha}$ & Amplitude of classical radiation\\
$\alpha$ & Rescaled amplitude\\
$C$ & Shorthand for $\sqrt{\frac{1}{2\omega V\epsilon_0}}$\\
$V$ & Quantization volume\\
$\bm{\epsilon}$ & Polarization vector\\
$\mathbf{k}$ & Wave vector\\
$\omega_L$ & Frequency of optical mode\\
$\omega_A$ & Energy difference between atomic ground and excited states\\
$\mathbf{D}$ & Atomic dipole operator\\
$\overline{D}$ & Shorthand for $\braket{g, \frac12||D||e, \frac12}$\\
$\Delta$ & Detuning $\omega_L - \omega_A$\\
$X$ & Random variable formed as absolute value of difference of two geometric variables \\
$S$ & Sum of random variables distributed like $X$
\end{tabular}
\end{ruledtabular}
\end{table*}
\end{document}